\documentclass[final,3p,times,twocolumn]{elsarticle}

\usepackage{graphicx}	
\usepackage{amssymb}
\usepackage{amsmath}	
\usepackage[dvipsnames]{xcolor}    
\usepackage[super]{nth}
\usepackage{microtype}
\usepackage{physics}
\usepackage{hyperref}
\usepackage{cleveref} 

\makeatletter
\def\ps@pprintTitle{%
 \let\@oddhead\@empty
 \let\@evenhead\@empty
 \let\@evenfoot\@oddfoot}
\makeatother

\newcommand*{\LCDM}{$\Lambda$CDM}

\newcommand*{\kmsMpc}{kms$^{-1}$Mpc$^{-1}$}

\newcommand{\CAMB}{\texttt{CAMB}}
\newcommand{\CosmoMC}{\texttt{CosmoMC}}

\begin{document}

\begin{frontmatter}

\title{Latest evidence for a late time vacuum--geodesic CDM interaction}

\author[ICG]{Natalie B. Hogg \corref{cor1}}
\author{Marco Bruni$^{\rm a, b}$}
\author[ICG]{Robert Crittenden}
\author[IFT]{Matteo Martinelli}
\author[Leiden]{Simone Peirone}

\address[ICG]{Institute  of  Cosmology  and  Gravitation,  University  of  Portsmouth, Burnaby  Road,  Portsmouth  PO1  3FX, United Kingdom}
\address[INFN]{INFN Sezione di Trieste, Via Valerio 2, 34127 Trieste, Italy}
\address[IFT]{Instituto de F\'isica T\'eorica UAM-CSIC, Campus de Cantoblanco, E-28049 Madrid, Spain}
\address[Leiden]{Institute Lorentz, Leiden University, PO Box 9506, Leiden 2300 RA, The Netherlands }
\cortext[cor1]{Corresponding author: natalie.hogg@port.ac.uk}

\begin{abstract}
We perform a reconstruction of the coupling function between vacuum energy and geodesic cold dark matter using the latest observational data. We bin the interaction in seventeen redshift bins but use a correlation prior to prevent rapid, unphysical oscillations in the coupling function. This prior also serves to eliminate any dependence of the reconstruction on the binning method. We use two different forms of the correlation prior, finding that both give similar results for the reconstruction of the dark matter -- dark energy interaction. Calculating the Bayes factor for each case, we find no meaningful evidence for deviation from the null interacting case, i.e. $\Lambda$CDM, in our reconstruction.
\end{abstract}


\end{frontmatter}


\section{Introduction} \label{sec:intro}
One of the most pressing questions in modern cosmology concerns the true nature of dark energy: what is the physical driver of the accelerated expansion of the Universe? This phenomenon of accelerating expansion was first established by \cite{Riess1998} and \cite{Perlmutter1999} and is generally attributed to the existence of a positive cosmological constant, $\Lambda$, in our Universe. While  the cosmological constant is the simplest proposed source of the accelerated expansion,  it is widely agreed that it has long suffered from theoretical problems: namely, why its observed value and theoretically predicted value differ so greatly \citep{Adler1995, Weinberg1989, Martin2012}, and why it has only come to dominate over the other components in the Universe relatively recently \citep{Velten:2014nra}. However, some advocate that these are not problems of cosmology but of particle physics or quantum field theory -- or perhaps not real problems at all \cite{Bianchi:2010uw}.

In addition to these theoretical and philosophical issues, the ever-increasing improvements in observational cosmology have begun to reveal tensions in the values of cosmological parameters within the wider $\Lambda$CDM model, particularly between high and low redshift measurements of these quantities. The most striking of these tensions is in the value of $H_0$, the Hubble parameter at redshift zero, which essentially informs us of the rate at which the Universe is expanding.

A value of $H_0$ can be measured without assuming a cosmological model, using low redshift probes such as Type Ia supernovae which rely on the distance ladder to fix the distance--redshift relation. A recent example of such a measurement is $H_0 = 73.45 \pm 1.66$ \kmsMpc \ \citep{Riess2018}. The value of $H_0$ can also be calculated using information coming from the cosmic microwave background (CMB) -- the caveat being that a cosmological model must be adopted. In doing so, the Planck collaboration reports $H_0 = 67.4 \pm 0.5 $ \kmsMpc, when $\Lambda$CDM is assumed \citep{Aghanim:2018eyx}. This signals a tension of more than $4 \sigma$  between the two measurements \citep{Riess:2019cxk}. 

Some other measurements of $H_0$ that are independent of the distance ladder do not seem to relax the tension. For example, the H0LiCOW doubly imaged quasar measurement found $H_0 = 72.5^{+2.1}_{-2.3}$ \kmsMpc \ \citep{Birrer2018}, the LIGO gravitational wave measurement found $H_0 = 70^{+12.0}_{-8.0}$ \kmsMpc \ \citep{Abbott:2017xzu} and the very recent Megamaser Cosmology Project constraint was found to be $H_0 = 73.9 \pm 3.0$ \kmsMpc \ \cite{Pesce:2020xfe}. 

The inverse distance ladder approach used in \citep{Macaulay:2018fxi} found $H_0 = 67.77 \pm 1.30$ \kmsMpc. This method entails anchoring the distance ladder using the baryon acoustic oscillation (BAO) signal combined with the size of the sound horizon at the drag epoch $r_s(z_d)$, which is in turn obtained either through CMB measurements or Big Bang nucleosynthesis constraints on the baryon density $\Omega_b h^2$. Anchoring the distance ladder in this way rather than to Cepheid variable stars results in a value of $H_0$ in agreement with the Planck 2018 result quoted above.

Furthermore, another tension is becoming apparent in the value of $\sigma_8$. This parameter is a measure of the growth of cosmological perturbations and hence of the large scale structure formation. The tension in its measured values is also between high and low redshift probes \citep{Battye:2014qga, Douspis:2018xlj}. 

While it is entirely possible that these tensions are present simply due to systematic errors or noise in the data, we must also consider the possibility -- following the well-known aphorism in statistics that ``all models are wrong'' \citep{Box1976} -- that the $\Lambda$CDM model is simply incorrect.

In an attempt to extricate cosmology from this rather alarming predicament, many alternatives to the $\Lambda$CDM model have been proposed. Some are as simple as allowing vacuum energy to interact with cold dark matter (CDM) (see e.g. \cite{Wands:2012vg, Salvatelli2014, Sola:2017lxc, Kumar:2017dnp, Martinelli:2019dau, Pan2019, Pan:2020zza}), others introduce an additional scalar field to drive the accelerated expansion (see \cite{Copeland2006} for a comprehensive review) and still others eliminate general relativity entirely, exploring modified gravity theories in which self-acceleration can be achieved (see e.g. \cite{Clifton2012, Joyce2016, Ezquiaga2017, Frusciante:2019xia} for details of various modified gravity models past and present). 

It is worth noting that some modified gravity models are motivated by the need to explain various features of large scale structure formation and thus do not necessarily alter the background cosmology. Since $H_0$ is a probe of the background expansion, models that do have some effect on the background cosmology are naturally more attractive when the motivation is to relax the $H_0$  tension.

In this work we choose the first option, introducing an interaction between the vacuum and cold dark matter, constructed in such a way that the cold dark matter remains geodesic, thus limiting any potentially pathological effects on structure growth. In a previous work \citep{Martinelli:2019dau}, we investigated whether a simple form of this interaction could relieve the tensions present in $\Lambda$CDM, testing the interaction acting in a single redshift bin and reconstructing the interaction using four redshift bins. We found that, while the interacting scenario does not manage to relieve cosmological tensions, it is not ruled out by current observational data.

In this work, we continue that investigation by increasing the number of redshift bins used in our reconstruction, thereby increasing the redshift range that the interaction acts over and ensuring a model-independent reconstruction. We also use the up-to-date Planck 2018 likelihood \cite{Aghanim:2018eyx}, instead of the 2015 likelihood used in our previous work. We study the constraining power of a theoretical prior acting across the bins and reconstruct the final interaction function. We then perform a Principal Component Analysis and calculate the Bayesian evidence for each case studied.

This paper is organised as follows: in Section \ref{sec:theory} we recapitulate the theory behind the interacting vacuum scenario that we test in this work. In Section \ref{sec:analysis} we describe the implementation and numerical analysis done, explaining the role of the theoretical priors and the reconstruction. We present our results and discussion in Section \ref{sec:results} and then conclude with Section \ref{sec:conclusions}.

\section{The interacting vacuum}\label{sec:theory}
In this section, we limit ourselves to a basic discussion of the interaction in a spatially flat Friedmann-Lema\^{i}tre-Robertson-Walker (FLRW) background. We note that the interaction is constructed so that CDM remains geodesic; in brief, this is because the energy-momentum flow 4-vector between the vacuum and cold dark matter, $Q^\mu$, can be projected in two parts, one parallel and one orthogonal to the CDM 4-velocity, $Q^\mu = Qu^\mu + f^\mu$, where $f^\mu$ is the momentum exchange, $f^\mu = a^\mu \rho_c$. 

We set this momentum exchange to be equal to zero, implying that the 4-acceleration $a^\mu$ must be zero and thus ensuring that CDM is always geodesic, as no additional acceleration due to the interaction acts on the CDM particles. With this choice, in the synchronous comoving gauge that we use to describe perturbations, the interaction is unperturbed and fully encoded in the background $Q$. 

We refer the reader to \cite{Wands:2012vg} and \cite{Martinelli:2019dau} for a detailed treatment of the covariant theory of the interaction, as well as the behaviour of linear perturbations in this theory and the effect of the interaction on structure growth. 

In a spatially flat FLRW background, the interaction is introduced between CDM and the vacuum in the following way:
\begin{align}
\dot{\rho}_c+ 3H \rho_c &= -Q, \label{eq:cdmcons} \\
\dot{V} &= Q, \label{eq:vaccons}
\end{align}
where $\rho_c$ and $V$ are the energy densities of CDM and the vacuum respectively, $H=\dot{a}/a$ is the Hubble expansion, with $a$ the cosmic scale factor,  and $Q$ is the energy exchange between the components. For $Q=0$, $ V$ is constant and we recover a cosmological constant, i.e.\ $\Lambda$CDM.

In order to reconstruct the behaviour of this interaction, we must choose a model  for $Q$. We make the following choice,
\begin{align}
    Q = -q H V,
\end{align}
so that the coupled energy conservation equations \eqref{eq:cdmcons}, \eqref{eq:vaccons} become
\begin{align}
\dot{\rho}_c+ 3H \rho_c &= q H V, \\
\dot{V} &= -q H V,
\end{align}
where $q=q(a)$ is a dimensionless function that encodes the strength of the coupling between CDM and the vacuum. A positive value of $q$ indicates that the vacuum is decaying and dark matter is growing, whereas a negative value of $q$ indicates that dark matter is decaying and the vacuum is growing. We aim to reconstruct the coupling as a function of redshift $z$, i.e.\ $q(z)$, using a cubic spline interpolation and a Gaussian process.

\section{Method}\label{sec:analysis}
In this section, we describe the numerical codes used and the modifications made to those codes, as well as the theoretical priors and data considered in our analysis. 

\subsection{Modifying \CAMB~and \CosmoMC}\label{subsec:codes}
The first step in our analysis is to constrain the coupling strength $q(a)$ with cosmological data. To this end we make use of modified versions of the \CAMB~\citep{Lewis:1999bs,Howlett:2012mh} and \CosmoMC~codes \citep{Lewis:2002ah, Lewis:2013hha}. We bin the interaction function $q(a)$ in terms of the cosmic scale factor, with $q_i$ being the  constant parameter value within the $i^{\rm th}$ bin. 

We choose to extend our previous four bin analysis presented in \cite{Martinelli:2019dau} to seventeen bins, with $i=1,...,17$; sixteen that are uniform in scale factor from $a=1.0$ to $a=0.14$, plus a single large bin that extends to $a\approx 0.0001$. We use \CosmoMC~to produce MCMC samples from the posterior distribution of the interaction parameter in each bin, plus the baryon and cold dark matter densities $\Omega_b h^2$ and $\Omega_c h^2$, the amplitude of the primordial power spectrum and the spectral index $A_s$ and  $n_s$, and the value of the Hubble parameter today, $H_0$. We use flat priors on these parameters, with the ranges specified in Table \ref{tab:priors}.

\begin{table}
\centering
\begin{tabular}{|c|c|}
\hline
Parameter          & Prior \\
\hline
$\Omega_bh^2$      &  $[0.005,0.1]$\\
$\Omega_ch^2$      &  $[0.001,0.99]$ \\
$H_0$              &  $[50,100]$\\
$\log{10^{10}A_s}$ &  $[2.0,4.0]$\\
$n_s$              &  $[0.8,1.2]$\\
$q_i$      &  $[-6.0,3.0]$  \\
\hline
\end{tabular}
\caption{Prior ranges of the parameters sampled in our analysis.}\label{tab:priors}
\end{table}

\subsection{Correlation prior}\label{subsec:cp}
Although we have no theoretically motivated model for the behaviour of the coupling as a function of scale factor (or, equivalently, time) we do have one theoretical prejudice: we do not expect the coupling function to oscillate rapidly, as we consider very fast changes of sign in the coupling function to be unphysical. We therefore take the step of including a theoretical prior on the coupling parameter that actively suppresses high frequency oscillations, thereby allowing the low frequency modes that are potentially present in the data to dominate. 

The theoretical prior takes the form of a scale-factor-dependent correlation between the values of the coupling function in each bin. Values of the function in neighbouring bins are correlated, with the correlation growing weaker for bins of greater separation. This \textit{correlation prior} was first proposed in \cite{Crittenden:2005wj} and the method has been subsequently used in the reconstruction of the dark energy equation of state function $w(z)$ by \cite{Crittenden2012, Wang:2018fng,Gerardi:2019obr}. The correlation prior method was also used by \cite{Dam:2019prv} to reconstruct the vacuum energy -- CDM interaction at low redshifts only, up to $z=1.5$.

The correlation prior has further benefits in addition to suppressing high frequency oscillations. It tends to improve the convergence speed of MCMC chains, as the correlation can help to constrain the coupling parameter in bins where the data is sparse. Reconstruction bias, i.e.\ the dependence of results on the binning strategy chosen is also controlled by the prior, provided that the number of bins is sufficiently large, as we will describe below.

Following \cite{Crittenden:2005wj}, we assume a correlation function that describes fluctuations around some fiducial model,
\begin{align}
&\xi(|a-a'|) \equiv \langle[q(a) - \bar{q}(a)] [q(a') - \bar{q}(a')] \rangle,\label{eq:corrfunc}
\intertext{and given a functional form for $\xi$, the corresponding covariance matrix can be found:}
&C_{ij} = \frac{1}{\Delta^2} \int^{a_i + \Delta}_{a_i} da \int^{a_j + \Delta}_{a_j} da' \xi(|a -a'|), \label{eq:covmat}
\end{align}
where $\Delta$ is the bin width, $\bar{q}$ is the fiducial model and $a$ is the cosmic scale factor. The fiducial model can be set to \LCDM \ (i.e. $\bar{q}(a_i) = 0$), but this may introduce an unwanted bias in favour of this model into our results, so for comparison we consider a case in which the fiducial model for each bin is calculated as the mean of that bin with its two neighbouring bins. We refer to these cases as \textit{fixed fiducial} and \textit{mean fiducial} respectively.

We use the Crittenden-Pogosian-Zhao (CPZ) form for the correlation function, as proposed in \cite{Crittenden:2005wj},
\begin{align}
 \xi(|a-a'|) = \xi(0)/ [1 + (|a-a'|/a_c)^2], 
\end{align}
where $a_c$ is the correlation length. As previously stated, we want to ensure that our results are independent of the number of bins used. To ensure that we eliminate this potential reconstruction bias, we require that
\begin{equation}
    N > N_{\rm eff}, \label{eq:neff}
\end{equation}

\noindent where $N$ is the number of bins, and 
\begin{equation}
    N_{\rm eff} = (a_{\rm max} - a_{\rm min}) / a_c. \label{eq:neff_two}
\end{equation}
The parameters $a_{\rm max}$ and $a_{\rm min}$ are the limits of the scale factor range used in our analysis, $a=1.0$ and $a=0.0001$. Following the previous results of \cite{Wang:2015wga, Wang:2014xca}, we choose $a_c = 0.06$. This means that $N_{\rm eff} = 16.7$. Therefore, to ensure that $N > N_{\rm eff}$, we choose $N=17$. 

The strength of the prior is determined by $\xi(0)$, but following \cite{Wang:2015wga}, we use the variance of the mean instead, defined as $\sigma^2_{q} \approx \pi \xi(0) a_c / (a_{\rm max}- a_{\rm min})$. We set $\sigma_q = 0.6$. We found that this choice is sufficient for the prior to provide some constraining power, but not so much that it completely dominates over the constraints from data in each bin. We discuss this point further in subsection \ref{subsec:PCA}.

\subsection{Observational data} \label{subsec:data}
The data used in this work is a combination of the Planck 2018 measurements of the CMB temperature and polarisation \citep{Aghanim:2018eyx}, the BAO measurements from the 6dF Galaxy Survey \citep{Beutler2011} and the combined  BAO and redshift space distortion (RSD) data from the SDSS DR12 consensus release \citep{Alam:2016hwk}, together with the Pantheon Type Ia supernovae sample \citep{Scolnic:2017caz}. 

We note that some works in the literature that find a resolution to the $H_0$ tension in an interacting dark energy scenario do so by omitting the BAO data from their analyses (see e.g. \cite{diValentino2019}). This is because, without using BAO, the high redshift constraint on $H_0$ becomes weaker, and a late time solution to the tension is possible. If BAO are used in combination with supernovae catalogues then late time solutions become disfavoured, and interacting dark energy models will therefore struggle to resolve the tension (see e.g. \cite{Poulin:2018zxs,Martinelli:2019krf}). However, this reasoning does not justify the exclusion of these datasets from model constraining analysis and we therefore make a point of including multiple BAO measurements in this work. 

We also note that, due to the coupling between the vacuum and cold dark matter in this scenario, RSD do not directly constrain the growth factor $f$ as they do in $\Lambda$CDM \cite{Martinelli:2019krf, Borges:2017jvi}. Instead, the RSD constrain what we denote as the interaction growth factor, $f_i$,
\begin{equation}
    f_i = f - \frac{Q}{H \rho_c},
\end{equation}
where $f$ is the usual growth factor for CDM, 
\begin{equation}
    f = \frac{d \ln D}{d \ln a},
\end{equation}
with D being the amplitude of the linear growing mode.

\section{Results and discussion} \label{sec:results}
In this section, we describe and discuss the main results of our investigation, beginning with the results of the MCMC analysis, then moving to the reconstruction of the coupling function, the Principal Component Analysis performed and finally the findings of our Bayesian evidence calculation.

\subsection{MCMC parameter inference} \label{subsec:MCMCresults}
\begin{figure*}
    \centering
    \includegraphics[width=0.9\textwidth]{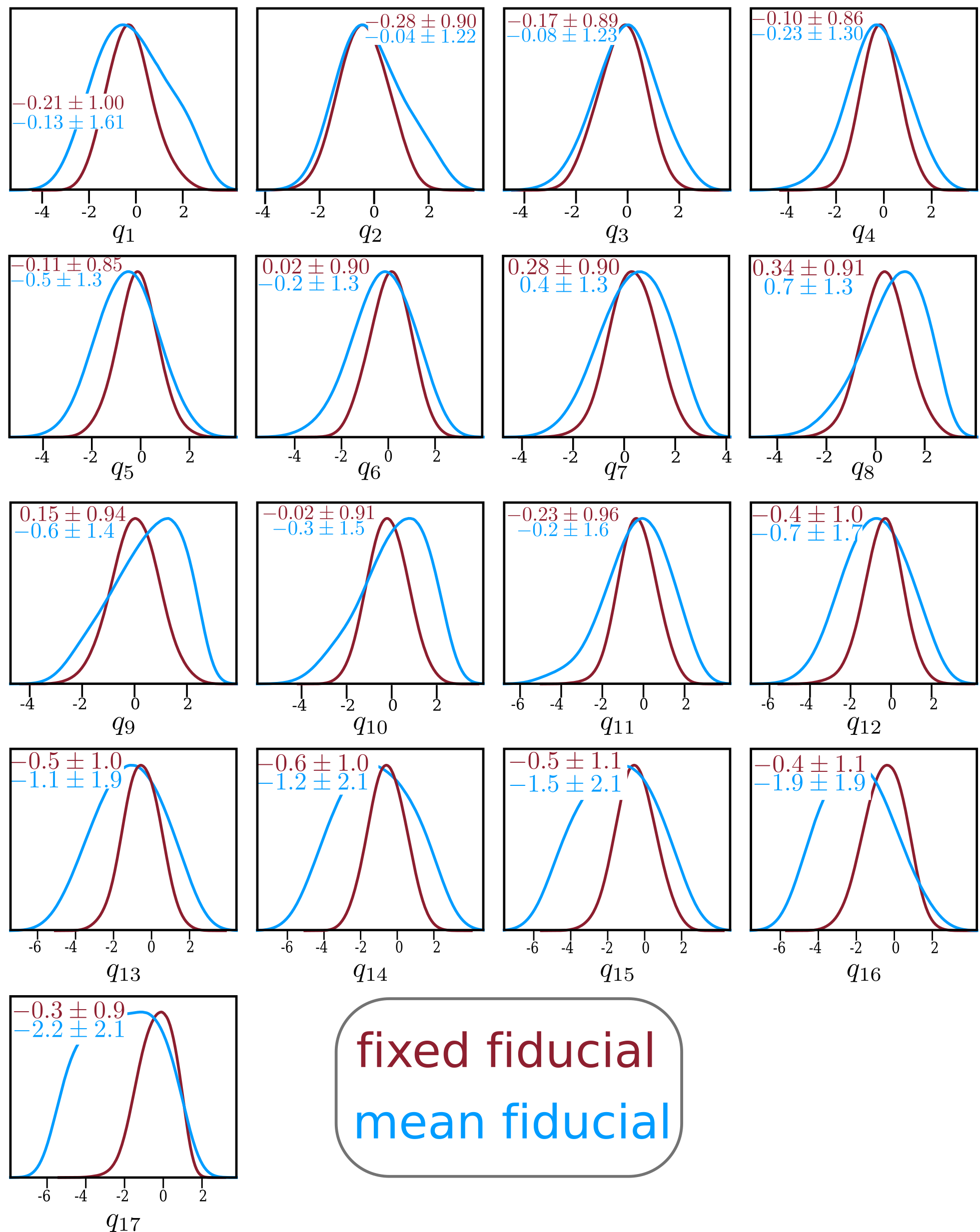}
    \caption{The 1D marginalised posteriors of the interaction parameter in each bin. In each panel we report the best fit value of the interaction parameters and their $68\%$ confidence level bounds for the fixed fiducial (red) and mean fiducial (blue) case. \label{fig:plot_qbins}}
\end{figure*}

In Figure \ref{fig:plot_qbins} we plot the 1D marginalised posteriors for the interaction parameter $q_i$ in each of the seventeen bins, where $i=1$ denotes the bin starting at $z=0$, up to $i=17$  for the wide bin at high redshift. The posterior distributions for $q_i$ are generally broader in the mean fiducial case compared to the fixed fiducial case. This is to be expected, as the mean fiducial case essentially has one additional free parameter with respect to the fixed fiducial, this being $\bar{q}$, the fiducial value for the correlation prior.

We find that the null interacting case ($q=0$), coinciding with the $\Lambda$CDM limit of the model, is always within $1\sigma$ of the achieved constraints. However, the bounds found on the interaction parameter in every bin means the interacting scenario is still viable. It is clear from an Ockham's razor standpoint that the $\Lambda$CDM scenario should be favoured over both the interacting cases. We quantify this statement using the Bayes factor in subsection \ref{subsec:evidence}.

\begin{figure*}
    \centering
    \includegraphics[width=0.49\textwidth]{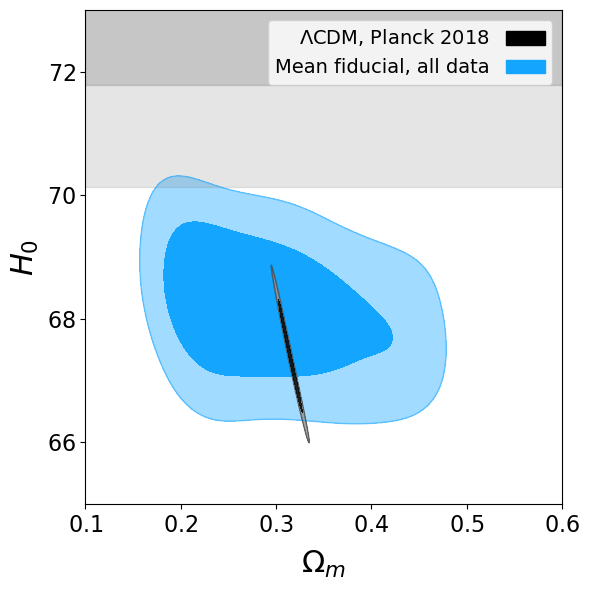}
    \includegraphics[width=0.49\textwidth]{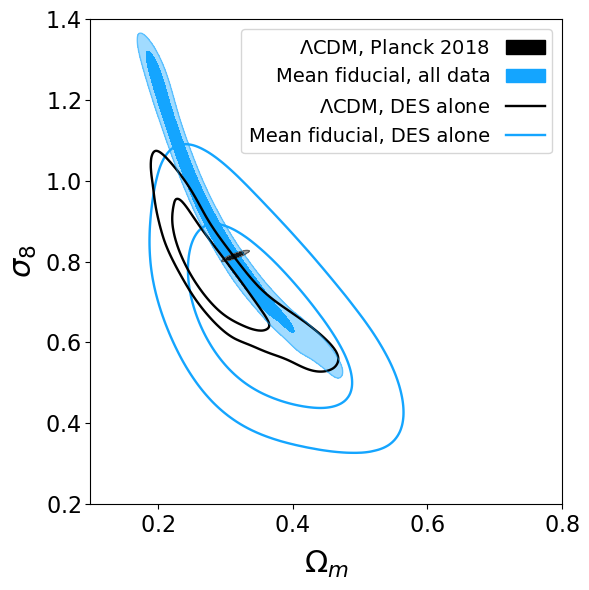}
    \caption{$68\%$ and $95\%$ confidence levels in the $H_0$ -- $\Omega_m$ plane (left panel) and $\Omega_m$ --  $\sigma_8$ plane (right panel) for the mean fiducial case. The grey bands in the left panel denote the $68\%$ and $95\%$ confidence levels of the Riess et al. local measurement of $H_0 = 73.45 \pm 1.66$ \kmsMpc \ \cite{Riess2018}.\label{fig:cosmo}}
\end{figure*}

Table \ref{tab:cosmoresults} shows the marginalised values of the standard cosmological parameters sampled in our MCMC analysis, while in Figure~\ref{fig:cosmo} we show the $2$D marginalised joint distributions for the cosmological parameters H$_0$, $\Omega_m$ (the total matter density parameter) and $\sigma_8$. To preserve the readability of the plot, we choose to only show the results of the mean fiducial case in this figure. As can be inferred from Figure \ref{fig:plot_qbins}, the constraints on the cosmological parameters in the fixed fiducial case are almost identical to those in the mean fiducial case. In both cases we found the value of $H_0$ to be completely consistent with the Planck 2018 $\Lambda$CDM value of $67.4 \pm 0.5$ \kmsMpc \citep{Aghanim:2018eyx}. The value of $\sigma_8$ given by Planck is $0.81 \pm 0.006$, which is comfortably within $1\sigma$ of the values for $\sigma_8$ we find in both interacting cases.

As discussed in the introduction, the tensions in the values of $H_0$ and $\sigma_8$ are commonly used as motivations for alternative models of dark energy. However, as we also found in our previous work \cite{Martinelli:2019dau}, the interacting vacuum fails to resolve the tensions when using the particular datasets chosen here. This can clearly be seen in the left panel of Figure \ref{fig:cosmo}, where the constraint on $H_0$ in the interacting scenario is shown in conjunction with both the Planck and local measurements. As we mentioned in subsection 3.3, for the case of the $H_0$ tension in particular, this is attributable to the fact that by including BAO and Type Ia supernovae in the same analysis the tension is shifted to a discrepancy in the sound horizon scale that cannot be resolved with a late time solution \cite{Aylor:2018drw,Arendse:2019hev,Knox:2019rjx}.

The situation is slightly less clear with respect to the $\sigma_8$ tension. In $\Lambda$CDM, the tension appears between CMB measurements coming from Planck and large scale structure constraints on growth such as those from the Dark Energy Survey (DES) \cite{Abbott:2017wau}. This mild tension can be seen in the right panel of Figure \ref{fig:cosmo}, with the $\Lambda$CDM constraints plotted in black, the filled contour corresponding to Planck and the open contour to DES. The DES constraint in the interacting scenario is plotted in the open blue contour -- again, in the interests of legibility we only show the mean fiducial case.

From this plot, we can see that the tension is relaxed in the interacting case, but only due to the increased size of the contours, which in turn is due to the additional free parameters in the interacting model with respect to $\Lambda$CDM. This should not be regarded as a true relaxation of the tension. Note that for the DES constraints presented here we implemented an aggressive cut of the non-linear scales in the data. Since we have no understanding of the non-linear regime in the interacting scenario we should not use this part of the data to obtain our constraints.

\begin{table*}
\centering
\begin{tabular}{|c|l|l|}
\hline
Parameter &  Fixed fiducial & Mean fiducial\\
\hline
$\Omega_b h^2$   & $0.022 \pm 0.00015$ & $0.022 \pm 0.00013$ \\
$\Omega_c h^2$   & $0.11  \pm 0.025$   & $0.11 \pm 0.030$ \\
$\log 10^{10} A$ & $3.05  \pm 0.0058$  & $3.05  \pm	0.0064$ \\
$n_s$            & $0.97  \pm 0.0040$  & $0.97 \pm 0.0048$\\
$H_0$            & $68.22 \pm 0.74$    & $68.15  \pm 0.80$\\
$\sigma_8$       & $0.91  \pm 0.18$    & $0.91   \pm 0.22$\\
\hline
\end{tabular}
\caption{Marginalised values of the cosmological parameters and their $68\%$ confidence level bounds. }\label{tab:cosmoresults}
\end{table*}

\subsection{Reconstructing the coupling function} \label{subsec:recon}
With the results of our MCMC analysis, we can reconstruct the coupling as a function of redshift. We show the results of using two different methods for the reconstruction: a simple cubic spline interpolation and a Gaussian process. 

A Gaussian process is defined as a collection of random variables, any finite number of which have a joint Gaussian distribution \citep{Rasmussen}. It is completely specified by its mean and its covariance. In practice, the random variables represent the value of a given function $f(x)$ at a location $x$. There are a wide range of choices for the covariance function, or kernel, that is used to relate the function values at each point. In this work, we choose to use one of the simplest, the squared exponential kernel, given by
\begin{align}
k(x, \tilde{x}) = \sigma^2 \exp\left(-\frac{(x - \tilde{x})}{2 \ell^2}\right).
\end{align}
The hyperparameters $\ell$ and $\sigma$ that appear in this kernel correspond to the approximate length scale over which the function varies, and the variance of the function at each point respectively. We optimise these by maximising the log-likelihood of the functions they produce. 

In summary, the Gaussian process takes some given training data and constructs the best possible function that describes that data, given the kernel imposed. The training data passed to the Gaussian process in our case are the mean posterior values of the coupling parameter in each bin along with the corresponding $1 \sigma$ errors given by our MCMC analysis, thereby allowing us to reconstruct the coupling function $q(z)$. 

There are many packages and codes available to perform Gaussian process regression. In this work, we use the Gaussian process regressor available in the Python library \texttt{george}\footnote{https://github.com/dfm/george}
\citep{Ambikasaran2015}. 

\begin{figure*}
    \centering
    \includegraphics[width=\textwidth]{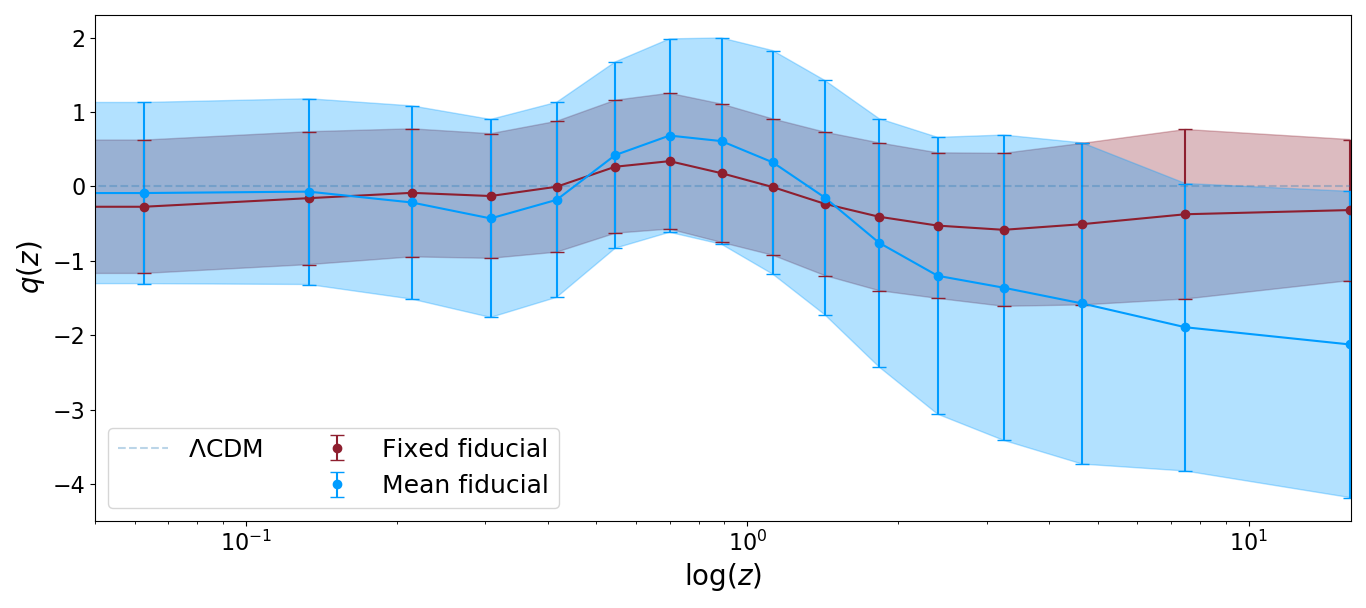}
    \caption{The results of the cubic spline reconstruction of the coupling function $q(z)$. Red and blue lines and areas refer to the fixed fiducial and mean fiducial cases respectively, and the shaded areas denote the $1 \sigma$ confidence interval. \label{fig:spline}}
\end{figure*}

\begin{figure*}
    \centering
    \includegraphics[width=\textwidth]{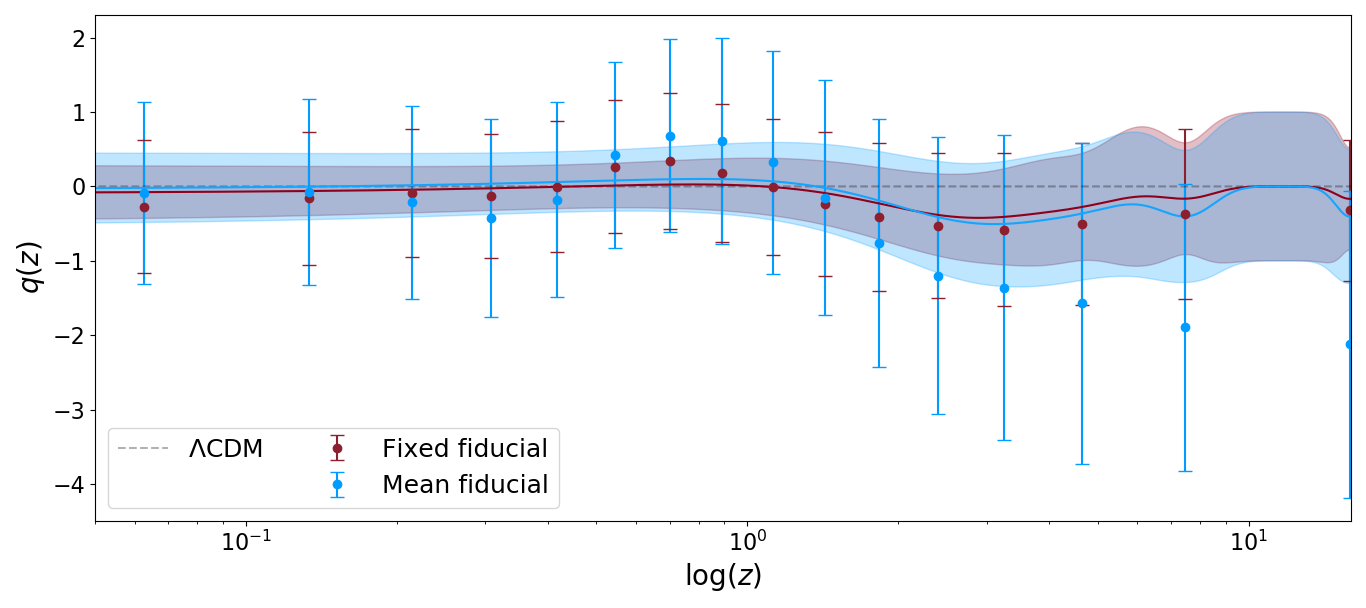}
    \caption{The results of the Gaussian process reconstruction of the coupling function $q(z)$. Red and blue lines and areas refer to the fixed fiducial and mean fiducial cases respectively, and the shaded areas denote the $1 \sigma$ confidence interval.  \label{fig:gp}}
\end{figure*}

The results of our reconstructions for the cubic spline and the Gaussian process are shown in \Cref{fig:spline,fig:gp} respectively. It is clear to see that the Gaussian process results in a smoother $q(z)$ function, but that the high redshift part of the reconstruction is biased towards the $\Lambda$CDM value of $q=0$, due to the baseline that the Gaussian process is fixed to return to in the absence of information. 

This is particularly obvious in the mean fiducial case, where the values of $q$ themselves are very negative but the combination of the Gaussian process baseline and the large $1\sigma$ errors on $q$ result in the reconstruction returning to zero. This is a problem that the cubic spline does not suffer from, hence the indication of a trend away from $\Lambda$CDM at high redshift in the mean fiducial case.

The most interesting features of the reconstruction are the points where $q(z)$ appears to peak or trough, for example, the peak at around $z=1$, which is clear in both the spline and Gaussian process, or the trough at around $z=3$, more obvious in the Gaussian process reconstruction. A promising line of enquiry would be to focus on the behaviour of the interaction  at these points by using additional datasets in the analysis, but as $z=3$ is beyond the upper limit of the commonly used low-redshift probes, such as Type Ia supernovae, exploring the interaction in detail at this epoch may be more difficult. 

A potential future constraint may come from the weak lensing of the Lyman-$\alpha$ forest in the spectra of high-redshift quasars, which probes the matter distribution at redshifts of 2 to 3.5 \citep{Croft:2017tur}. Furthermore, the Square Kilometre Array is predicted to be able to probe redshifts of between 3 and 25 using 21cm intensity mapping \citep{Braun2015, SKA2018}. Both of these new techniques could therefore be used to constrain any interacting dark energy model which affects large scale structure growth or has other high redshift effects. 

\subsection{Principal Component Analysis} \label{subsec:PCA}
In this work, we have aimed to be agnostic when it comes to the reconstruction of the interaction function and so used a larger number of bins than in \cite{Martinelli:2019dau}, i.e.\ the minimum number to satisfy the criterion given by equation \eqref{eq:neff}. However, it is also possible to investigate how many modes in the result are informed by the observational data used and whether any are informed by the prior alone, and thus understand how many effective additional degrees of freedom our reconstruction has \citep{Huterer:2002hy}. To do this, we perform a principal component analysis.

Principal component analysis, or PCA, can be thought of as finding the directions in the data that carry the most information. It also acts to decorrelate the errors on the interaction parameter in each bin. In practice, this involves computing the eigenvalues and eigenvectors of the inverse covariance matrix (i.e. the Fisher matrix) of the data. In our case, the covariance matrix is one of the products obtained after running \texttt{GetDist}\footnote{https://github.com/cmbant/getdist}\cite{Lewis:2019xzd} on our MCMC chains. We perform the PCA on the Fisher matrix for the $q_i$ alone, after marginalising over the other cosmological and nuisance parameters.

The Fisher matrix is given by
\begin{equation}
    F = W^T \Lambda W,
\end{equation}
where $W$ is the decorrelation matrix and its rows define the eigenvectors; $\Lambda$ is a diagonal matrix whose elements are the eigenvalues $\lambda_i$. The eigenvalues correspond to the amount of variance carried in each principal component and therefore determine how well $q_i$ can be measured, i.e. $\sigma(q_i) = \lambda_i^{-1/2}$.

After finding the eigenvalues and eigenvectors of the covariance matrix, the eigenvectors are sorted according to decreasing value of their corresponding eigenvalues. The first eigenvector after this sort is performed corresponds to the first principal component, the second eigenvector corresponds to the second principal component and so on, until the $N^{\rm th}$ eigenvector for the  $N^{\rm th}$ principal component is found (where the covariance matrix is $N \times N$).

\begin{figure}
    \centering
    \includegraphics[width=\columnwidth]{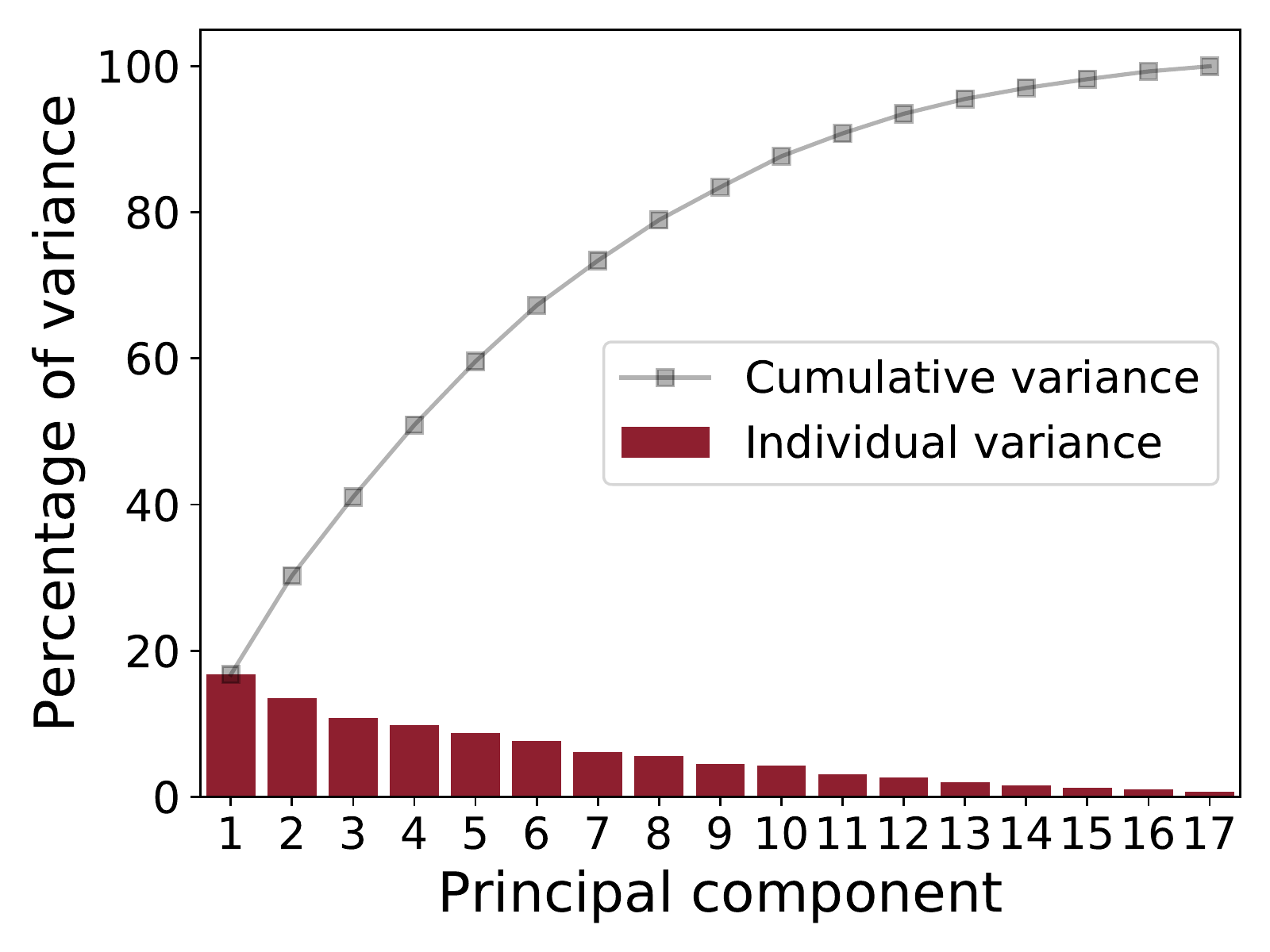}
    \caption{Percentage variance explained by each principal component in the fixed fiducial case. \label{fig:fixpca}}
\end{figure}

\begin{figure}
    \centering
    \includegraphics[width=\columnwidth]{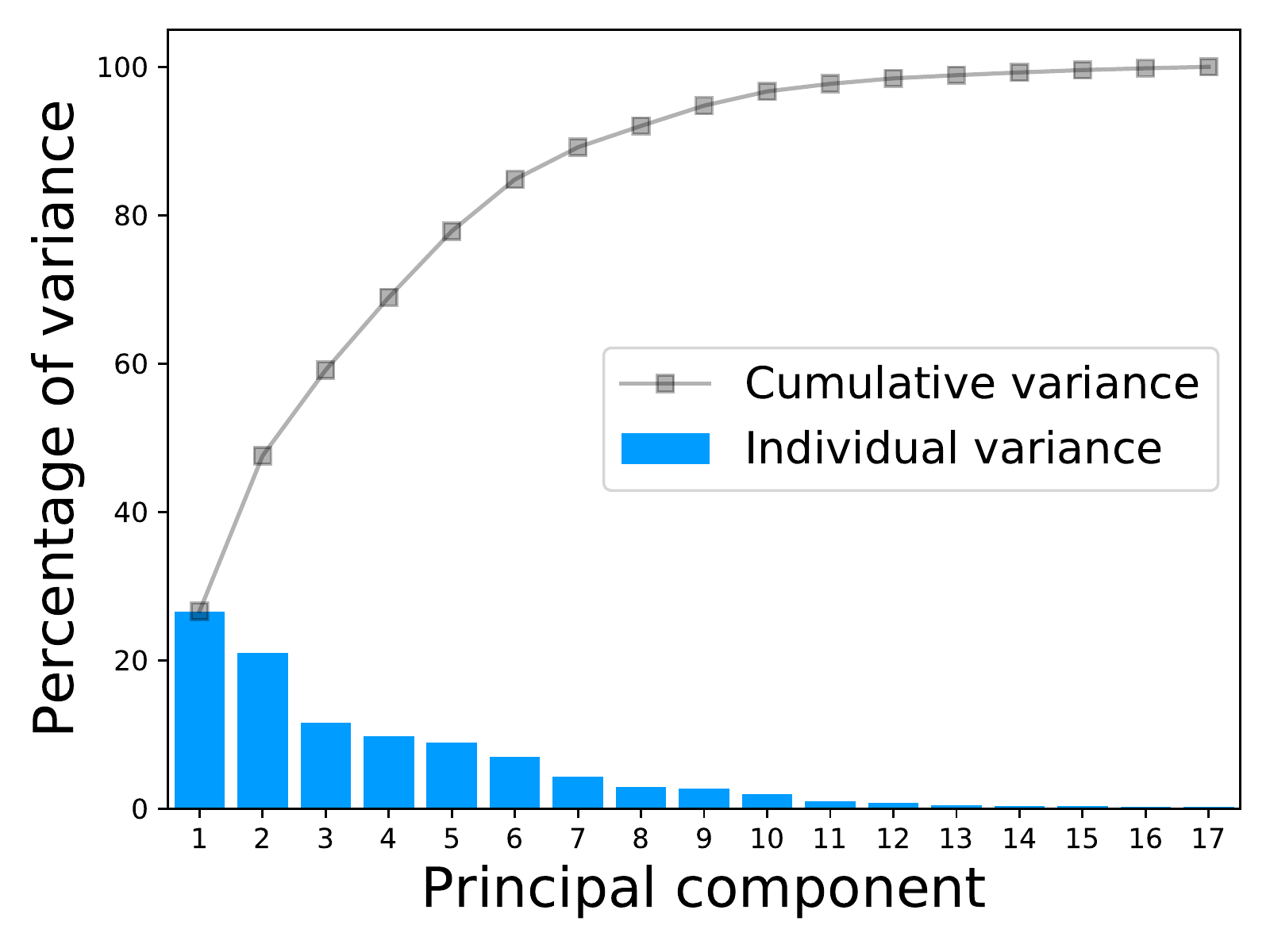}
    \caption{Percentage variance explained by each principal component in the mean fiducial case.\label{fig:meanpca}}
\end{figure}

We show the results of our PCA in \Cref{fig:fixpca,fig:meanpca}. From these plots we can see that in the fixed fiducial case around 15\% of the total variance is in the first principal component, we reach around 50\% with four principal components and 90\% with 10. These results indicate that it would be unwise to reduce the effective degrees of freedom by discarding some of the principal components, as even the higher components contain a significant amount of information (above PC10 the remaining seven components together still contain approximately 11\% of the variance). This is less true in the case of the mean fiducial, in which around 25\% of the total variance is contained in the first principal component, rising to nearly 50\% with just two principal components and reaching 90\% with seven. The final four principal components together contain just 1\% of the variance. 

\begin{figure}
    \centering
    \includegraphics[width=\columnwidth]{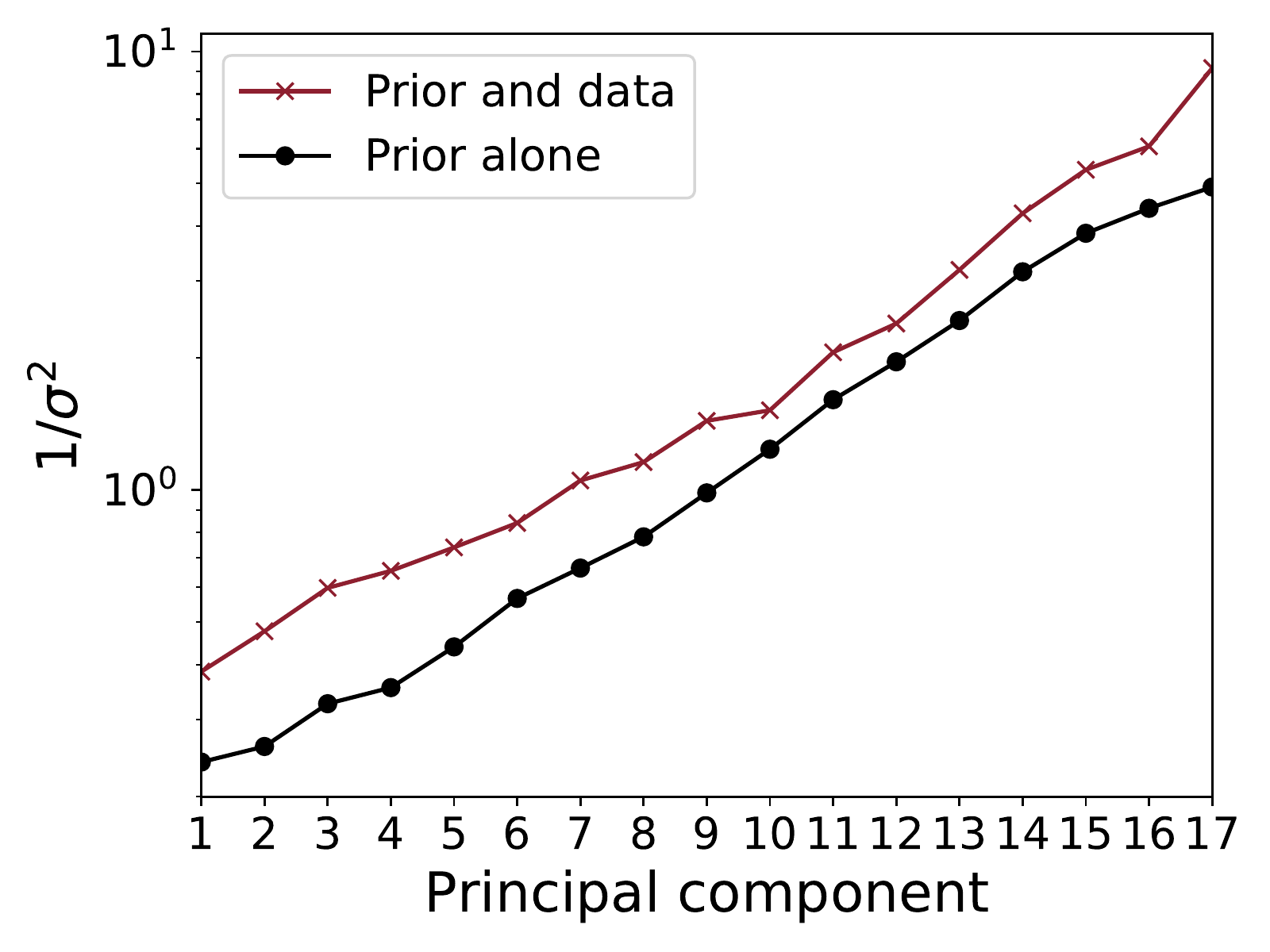}
    \caption{Showing that the data permeates all the modes. \label{fig:eigenvalues}}
\end{figure}

To investigate whether the correlation prior dominates over the data, we also ran an MCMC chain without any data, using the prior alone to constrain the interaction. This prior alone case used $\bar{q}=0$, as in the case of the fixed fiducial. We plot the eigenvalues of the fixed fiducial case and the prior alone case as a function of principal component number in \Cref{fig:eigenvalues}. This plot shows that the data permeates all the modes, meaning that the prior does not completely dominate over the data at any point and thus the selected prior strength was indeed sufficient to help constrain the interaction without washing out the information coming from the data. Note that we only show the result for the case of the fixed fiducial prior alone and the fixed fiducial prior plus data, as the result for the mean fiducial is extremely similar. 

If we had found that the data dominated for say, the first three principal components and then the prior dominated over the rest, we would be able to conclude that our analysis effectively only had an additional three degrees of freedom compared to the $\Lambda$CDM case. However, this does not equate to doing an analysis using only three bins, as the principal components do not correspond to the bins themselves, but to the eigenvectors of the covariance matrix of the interaction parameter in each bin. We therefore conclude that the best strategy for an analysis such as this is to use as many bins as is computationally feasible, with the correlation prior being used to help constrain bins where data is scarce. The alternative is to increase the strength of the correlation prior, but this comes with its own pitfalls, as if the prior is too strong, it will completely wash out any contribution from the data. A balance can be achieved, but to ensure that the reconstruction remains independent of the number of bins used, the correlation length and therefore the prior strength should be determined by following equations \eqref{eq:neff} and \eqref{eq:neff_two}.

\subsection{Bayesian evidence and $\chi^2$ } \label{subsec:evidence}
Finally, we want to compare the results for each case in a Bayesian way, which means making use of Bayes' theorem \citep{Bayes1763}:
\begin{align}
    P(\theta | D, M) &= \frac{P(D | \theta, M) P(\theta | M) }{P(D | M)}, \label{eq:bayes}
\end{align}
where $\theta$ is the parameter vector, $D$ is the data vector and $M$ is the model. The numerator contains the likelihood and the prior, and the denominator is the evidence (sometimes known as the marginal likelihood). These combine to form the posterior probability distribution $P(\theta | D, M)$, which is the distribution sampled in our MCMC analysis.

As noted by \cite{Linares2019}, the use of model selection criteria such as the Bayesian Information Criterion (BIC), Akaike Information Criterion (AIC) and Deviance Information Criterion (DIC) are not strictly Bayesian as they do not take into account the prior information. We therefore use the Bayes factor as our model comparison tool, defined in the following way: 
\begin{align}
\log B_{12} &= \log \left[\frac{P(D|M_1)}{P(D|M_2)}\right], \\ 
&= \log [P(D|M_1)] - \log [P(D|M_2)],
\end{align}
where $D$ is the data vector, $M_1$ and $M_2$ are the models to be compared, and $P(D|M)$ is the Bayesian evidence, the normalising factor in Bayes' theorem.

We calculate the Bayesian evidence from our MCMC chains for both of the two correlation prior cases studied to determine the support for each case over $\Lambda$CDM. This analysis was performed using the \texttt{MCEvidence} code as presented in \cite{Heavens:2017afc}. In each case, we use $\Lambda$CDM as model 1. We summarise our findings in Table \ref{tab:evidence}.

To interpret these values, we make use of the Jeffreys scale, as shown in Table \ref{tab:jeffreys}. As pointed out in \cite{Hobson}, the qualitative interpretations originally given by Jeffreys \cite{Jeffreys} are quite strong in the context of cosmology, where choosing suitable priors can often be an uncertain process. We therefore adopt the interpretations given in \cite{Hobson}. 

\begin{table*}
\centering
\begin{tabular}{|c|c|c|}
\hline
Case & Bayes factor ($\log B_{12}$)& $\Delta \chi^2$ \\
\hline
Fixed fiducial & 1.64 & -2.5\\
Mean fiducial &-0.52 &  -2.2\\
\hline
\end{tabular}
\caption{The Bayes factor and $\Delta \chi^2 $ for each case. \label{tab:evidence}}
\end{table*}

\begin{table*}
\centering
\begin{tabular}{|c|c|}
\hline
Bayes factor & Interpretation\\
\hline
$ \abs{\log B_{12}} < 1$ & Not worth more than a bare mention\\ 
$ 1 <  \abs{\log B_{12}} < 2.5 $ & Weak \\ 
$2.5 <  \abs{\log B_{12}} < 5$ & Significant \\ 
$5 <  \abs{\log B_{12}} $ & Strong \\ 
\hline
\end{tabular}
\caption{The Jeffreys scale, originally given in \protect\cite{Jeffreys} and modified in \protect\cite{Hobson}.\label{tab:jeffreys}}
\end{table*}

We find that the Bayes factor for the fixed fiducial case is 1.64. According to the Jeffreys scale, this reflects a weak preference for $\Lambda$CDM over the interacting case. The Bayes factor for the mean fiducial case is -0.52. In our evidence calculation, negative values indicate that model 2 is preferred over model 1, where model 1 is always $\Lambda$CDM. This result is therefore a slight indication for the mean fiducial case being favoured over $\Lambda$CDM. However, according to the Jeffreys scale, the very small absolute value of the Bayes factor means this is not worth more than a bare mention.

The fact that we find stronger evidence in favour of $\Lambda$CDM in the case where the fiducial is fixed as $q=0$ could point to a slight bias in the results caused by the choice of fiducial model. However, the evidence in favour of the interaction when the fiducial is calculated as the mean of neighbouring bins is sufficiently small for us to confidently say that the choice of fiducial model does not drastically alter the result of a reconstruction.

However, it has been argued that the Bayesian evidence is not a good model comparison tool when there is uncertainty in the choice of priors \cite{Efstathiou2008}. We therefore also compute the $\Delta \chi^2$ for each case, removing the contribution of the priors to the $\chi^2$ so that the values we compare come from the data only. We find $\Delta \chi^2 = -2.5$ in the fixed fiducial case and  $\Delta \chi^2 = -2.2$ in the mean fiducial case, neither of which represents a significant improvement in fit over $\Lambda$CDM. 

In summary, it is clear that we cannot conclusively state that $\Lambda$CDM is preferred over the interacting case, but the hints given by the evidence indicate that an interesting future direction would be to repeat this type of analysis with the newest datasets as they are released, to see if there is any strengthening in the evidence for or against $\Lambda$CDM. It is also worthwhile studying what possible improvements on current constraints can be made by future surveys. 

\section{Conclusions} \label{sec:conclusions}
In this work we have reconstructed a dark matter -- vacuum energy interaction, using a correlation prior to control the reconstruction bias. We implemented two different versions of the prior: a fiducial value for the prior that is fixed in each bin and a fiducial value that is computed as the mean of the neighbouring bins. 

In our model comparison, we found evidence in favour of $\Lambda$CDM over the fixed fiducial model, but the Bayes factor in that case was small enough to classify the evidence on the Jeffreys scale as weak. In contrast, we found evidence for an interaction when comparing the $\Lambda$CDM case to the mean fiducial case, but the Jeffreys scale in that case classified the evidence as not worth more than a bare mention.

From our work, it is clear that a correlation prior, when effectively tuned so as not to drown out the constraining power of the data, can improve the convergence speed of high-dimensionality MCMC sampling. The prior also eliminates any potential reconstruction bias, making it a good choice for any form of reconstructive analysis.

Finally, we note that many recent works have found evidence for an interaction in the dark sector or for dynamical dark energy (see e.g. \cite{Zhao:2017cud, Wang:2018fng, Yang:2018qmz, Yang:2018qec, Lucca:2020zjb}), and the attractive properties of such models combined with the deficiencies of $\Lambda$CDM that we discussed in the introduction are sufficient to merit their continued study. It is clear that the large amounts of new data which upcoming surveys are expected to yield will be a vital clue in the hunt for the true nature of dark energy, and robust forecasting for the constraints these surveys are expected to provide on alternative dark energy models will become ever-more important.

\section*{CRediT authorship contribution statement}
\textbf{Natalie B. Hogg:} Software, Formal analysis, Investigation, Writing -- original draft. \textbf{Marco Bruni:} Methodology, Conceptualization, Writing -- review \& editing, Supervision. \textbf{Robert Crittenden:} Methodology, Conceptualization, Writing -- review \& editing, Supervision. \textbf{Matteo Martinelli:} Methodology, Conceptualization, Writing -- review \& editing, Supervision. \textbf{Simone Peirone:} Software, Writing -- review \& editing, Visualization.

\section*{Declaration of competing interest}

The authors declare that they have no known competing financial interests or personal relationships that could have appeared to influence the work reported in this paper.

\section*{Acknowledgements}
We thank Minas Karamanis, David Wands, Yuting Wang and Gong-Bo Zhao for enlightening discussions. This paper is based upon work from COST action CA15117 (CANTATA), supported by COST (European Cooperation in Science and Technology). Numerical computations were done on the SCIAMA High Performance Computer (HPC) cluster which is supported by the ICG, SEPNet and the University of Portsmouth.  NBH is supported by UK STFC studentship ST/N504245/1. MB and RC are supported by UK STFC Grant No. ST/S000550/1. MM has received the support of a fellowship from ``la Caixa'' Foundation (ID 100010434), with fellowship code LCF/BQ/PI19/11690015, and the support of the Spanish Agencia Estatal de Investigacion through the grant “IFT Centro de Excelencia Severo Ochoa SEV-2016-0597”. SP acknowledges support from the NWO and the Dutch Ministry of Education, Culture and Science (OCW), and also from the D-ITP consortium, a programme of the NWO that is funded by the OCW.

\bibliographystyle{elsarticle-num-names}
\bibliography{17bins}

\begin{thebibliography}{68}
\expandafter\ifx\csname natexlab\endcsname\relax\def\natexlab#1{#1}\fi
\providecommand{\url}[1]{\texttt{#1}}
\providecommand{\href}[2]{#2}
\providecommand{\path}[1]{#1}
\providecommand{\DOIprefix}{doi:}
\providecommand{\ArXivprefix}{arXiv:}
\providecommand{\URLprefix}{URL: }
\providecommand{\Pubmedprefix}{pmid:}
\providecommand{\doi}[1]{\href{http://dx.doi.org/#1}{\path{#1}}}
\providecommand{\Pubmed}[1]{\href{pmid:#1}{\path{#1}}}
\providecommand{\bibinfo}[2]{#2}
\ifx\xfnm\relax \def\xfnm[#1]{\unskip,\space#1}\fi
\bibitem[{{Riess} et~al.(1998)}]{Riess1998}
\bibinfo{author}{A.~G. {Riess}}, et~al.,
\newblock \bibinfo{title}{{Observational Evidence from Supernovae for an
  Accelerating Universe and a Cosmological Constant}},
\newblock \bibinfo{journal}{The Astronomical Journal} \bibinfo{volume}{116}
  (\bibinfo{year}{1998}) \bibinfo{pages}{1009--1038}.
  \DOIprefix\doi{10.1086/300499}.
  \href{http://arxiv.org/abs/astro-ph/9805201}{{\tt arXiv:astro-ph/9805201}}.
\bibitem[{{Perlmutter} et~al.(1999)}]{Perlmutter1999}
\bibinfo{author}{S.~{Perlmutter}}, et~al.,
\newblock \bibinfo{title}{{Measurements of {$\Omega$} and {$\Lambda$} from 42
  High-Redshift Supernovae}},
\newblock \bibinfo{journal}{The Astrophysical Journal} \bibinfo{volume}{517}
  (\bibinfo{year}{1999}) \bibinfo{pages}{565--586}.
  \DOIprefix\doi{10.1086/307221}.
  \href{http://arxiv.org/abs/astro-ph/9812133}{{\tt arXiv:astro-ph/9812133}}.
\bibitem[{{Adler} et~al.(1995){Adler}, {Casey}, and {Jacob}}]{Adler1995}
\bibinfo{author}{R.~J. {Adler}}, \bibinfo{author}{B.~{Casey}},
  \bibinfo{author}{O.~C. {Jacob}},
\newblock \bibinfo{title}{{Vacuum catastrophe: An elementary exposition of the
  cosmological constant problem}},
\newblock \bibinfo{journal}{American Journal of Physics} \bibinfo{volume}{63}
  (\bibinfo{year}{1995}) \bibinfo{pages}{620--626}.
  \DOIprefix\doi{10.1119/1.17850}.
\bibitem[{Weinberg(1989)}]{Weinberg1989}
\bibinfo{author}{S.~Weinberg},
\newblock \bibinfo{title}{The cosmological constant problem},
\newblock \bibinfo{journal}{Reviews of Modern Physics} \bibinfo{volume}{61}
  (\bibinfo{year}{1989}) \bibinfo{pages}{1--23}.
\bibitem[{{Martin}(2012)}]{Martin2012}
\bibinfo{author}{J.~{Martin}},
\newblock \bibinfo{title}{{Everything you always wanted to know about the
  cosmological constant problem (but were afraid to ask)}},
\newblock \bibinfo{journal}{Comptes Rendus Physique} \bibinfo{volume}{13}
  (\bibinfo{year}{2012}) \bibinfo{pages}{566--665}.
  \DOIprefix\doi{10.1016/j.crhy.2012.04.008}.
  \href{http://arxiv.org/abs/1205.3365}{{\tt arXiv:1205.3365}}.
\bibitem[{Velten et~al.(2014)Velten, vom Marttens, and
  Zimdahl}]{Velten:2014nra}
\bibinfo{author}{H.~E.~S. Velten}, \bibinfo{author}{R.~F. vom Marttens},
  \bibinfo{author}{W.~Zimdahl},
\newblock \bibinfo{title}{{Aspects of the cosmological “coincidence
  problem”}},
\newblock \bibinfo{journal}{Eur. Phys. J.} \bibinfo{volume}{C74}
  (\bibinfo{year}{2014}) \bibinfo{pages}{3160}.
  \DOIprefix\doi{10.1140/epjc/s10052-014-3160-4}.
  \href{http://arxiv.org/abs/1410.2509}{{\tt arXiv:1410.2509}}.
\bibitem[{Bianchi and Rovelli(2010)}]{Bianchi:2010uw}
\bibinfo{author}{E.~Bianchi}, \bibinfo{author}{C.~Rovelli},
\newblock \bibinfo{title}{{Why all these prejudices against a constant?}}
  (\bibinfo{year}{2010}). \href{http://arxiv.org/abs/1002.3966}{{\tt
  arXiv:1002.3966}}.
\bibitem[{{Riess} et~al.(2018){Riess}, {Casertano}, {Yuan}, {Macri},
  {Anderson}, {MacKenty}, {Bowers}, {Clubb}, {Filippenko}, {Jones}, and
  {Tucker}}]{Riess2018}
\bibinfo{author}{A.~G. {Riess}}, \bibinfo{author}{S.~{Casertano}},
  \bibinfo{author}{W.~{Yuan}}, \bibinfo{author}{L.~{Macri}},
  \bibinfo{author}{J.~{Anderson}}, \bibinfo{author}{J.~W. {MacKenty}},
  \bibinfo{author}{J.~B. {Bowers}}, \bibinfo{author}{K.~I. {Clubb}},
  \bibinfo{author}{A.~V. {Filippenko}}, \bibinfo{author}{D.~O. {Jones}},
  \bibinfo{author}{B.~E. {Tucker}},
\newblock \bibinfo{title}{{New Parallaxes of Galactic Cepheids from Spatially
  Scanning the Hubble Space Telescope: Implications for the Hubble Constant}},
\newblock \bibinfo{journal}{The Astrophysical Journal} \bibinfo{volume}{855}
  (\bibinfo{year}{2018}) \bibinfo{pages}{136}.
  \DOIprefix\doi{10.3847/1538-4357/aaadb7}.
  \href{http://arxiv.org/abs/1801.01120}{{\tt arXiv:1801.01120}}.
\bibitem[{Aghanim et~al.(2018)}]{Aghanim:2018eyx}
\bibinfo{author}{N.~Aghanim}, et~al.,
\newblock \bibinfo{title}{{Planck 2018 results. VI. Cosmological parameters}},
\newblock \bibinfo{journal}{ArXiv e-prints}  (\bibinfo{year}{2018}).
  \href{http://arxiv.org/abs/1807.06209}{{\tt arXiv:1807.06209}}.
\bibitem[{{Riess} et~al.(2019){Riess}, {Casertano}, {Yuan}, {Macri}, and
  {Scolnic}}]{Riess:2019cxk}
\bibinfo{author}{A.~G. {Riess}}, \bibinfo{author}{S.~{Casertano}},
  \bibinfo{author}{W.~{Yuan}}, \bibinfo{author}{L.~M. {Macri}},
  \bibinfo{author}{D.~{Scolnic}},
\newblock \bibinfo{title}{{Large Magellanic Cloud Cepheid Standards Provide a
  1\% Foundation for the Determination of the Hubble Constant and Stronger
  Evidence for Physics beyond {\ensuremath{\Lambda}}CDM}},
\newblock \bibinfo{journal}{The Astrophysical Journal} \bibinfo{volume}{876}
  (\bibinfo{year}{2019}) \bibinfo{pages}{85}.
  \DOIprefix\doi{10.3847/1538-4357/ab1422}.
  \href{http://arxiv.org/abs/1903.07603}{{\tt arXiv:1903.07603}}.
\bibitem[{{Birrer} et~al.(2019){Birrer}, {Treu}, {Rusu}, {Bonvin}, {Fassnacht},
  {Chan}, {Agnello}, {Shajib}, {Chen}, and {Auger}}]{Birrer2018}
\bibinfo{author}{S.~{Birrer}}, \bibinfo{author}{T.~{Treu}},
  \bibinfo{author}{C.~E. {Rusu}}, \bibinfo{author}{V.~{Bonvin}},
  \bibinfo{author}{C.~D. {Fassnacht}}, \bibinfo{author}{J.~H.~H. {Chan}},
  \bibinfo{author}{A.~{Agnello}}, \bibinfo{author}{A.~J. {Shajib}},
  \bibinfo{author}{G.~C.~F. {Chen}}, \bibinfo{author}{M.~{Auger}},
\newblock \bibinfo{title}{{H0LiCOW - IX. Cosmographic analysis of the doubly
  imaged quasar SDSS 1206+4332 and a new measurement of the Hubble constant}},
\newblock \bibinfo{journal}{Monthly Notices of the Royal Astronomical Society}
  \bibinfo{volume}{484} (\bibinfo{year}{2019}) \bibinfo{pages}{4726--4753}.
  \DOIprefix\doi{10.1093/mnras/stz200}.
  \href{http://arxiv.org/abs/1809.01274}{{\tt arXiv:1809.01274}}.
\bibitem[{Abbott et~al.(2017)}]{Abbott:2017xzu}
\bibinfo{author}{B.~P. Abbott}, et~al. (\bibinfo{collaboration}{LIGO
  Scientific, VINROUGE, Las Cumbres Observatory, DES, DLT40, Virgo, 1M2H, Dark
  Energy Camera GW-E, MASTER}),
\newblock \bibinfo{title}{{A gravitational-wave standard siren measurement of
  the Hubble constant}},
\newblock \bibinfo{journal}{Nature} \bibinfo{volume}{551}
  (\bibinfo{year}{2017}) \bibinfo{pages}{85--88}.
  \DOIprefix\doi{10.1038/nature24471}.
  \href{http://arxiv.org/abs/1710.05835}{{\tt arXiv:1710.05835}}.
\bibitem[{Pesce et~al.(2020)}]{Pesce:2020xfe}
\bibinfo{author}{D.~W. Pesce}, et~al.,
\newblock \bibinfo{title}{{The Megamaser Cosmology Project. XIII. Combined
  Hubble constant constraints}}  (\bibinfo{year}{2020}).
  \href{http://arxiv.org/abs/2001.09213}{{\tt arXiv:2001.09213}}.
\bibitem[{{Macaulay} et~al.(2019){Macaulay}, {Nichol}, {Bacon}, {Brout},
  {Davis}, {Zhang}, {Bassett}, {Scolnic}, {M{\"o}ller}, and
  {D'Andrea}}]{Macaulay:2018fxi}
\bibinfo{author}{E.~{Macaulay}}, \bibinfo{author}{R.~C. {Nichol}},
  \bibinfo{author}{D.~{Bacon}}, \bibinfo{author}{D.~{Brout}},
  \bibinfo{author}{T.~M. {Davis}}, \bibinfo{author}{B.~{Zhang}},
  \bibinfo{author}{B.~A. {Bassett}}, \bibinfo{author}{D.~{Scolnic}},
  \bibinfo{author}{A.~{M{\"o}ller}}, \bibinfo{author}{C.~B. {D'Andrea}},
\newblock \bibinfo{title}{{First cosmological results using Type Ia supernovae
  from the Dark Energy Survey: measurement of the Hubble constant}},
\newblock \bibinfo{journal}{Monthly Notices of the Royal Astronomical Society}
  \bibinfo{volume}{486} (\bibinfo{year}{2019}) \bibinfo{pages}{2184--2196}.
  \DOIprefix\doi{10.1093/mnras/stz978}.
  \href{http://arxiv.org/abs/1811.02376}{{\tt arXiv:1811.02376}}.
\bibitem[{Battye et~al.(2015)Battye, Charnock, and Moss}]{Battye:2014qga}
\bibinfo{author}{R.~A. Battye}, \bibinfo{author}{T.~Charnock},
  \bibinfo{author}{A.~Moss},
\newblock \bibinfo{title}{{Tension between the power spectrum of density
  perturbations measured on large and small scales}},
\newblock \bibinfo{journal}{Physical Review D} \bibinfo{volume}{91}
  (\bibinfo{year}{2015}) \bibinfo{pages}{103508}.
  \DOIprefix\doi{10.1103/PhysRevD.91.103508}.
  \href{http://arxiv.org/abs/1409.2769}{{\tt arXiv:1409.2769}}.
\bibitem[{Douspis et~al.(2018)Douspis, Salvati, and Aghanim}]{Douspis:2018xlj}
\bibinfo{author}{M.~Douspis}, \bibinfo{author}{L.~Salvati},
  \bibinfo{author}{N.~Aghanim},
\newblock \bibinfo{title}{{On the Tension between Large Scale Structures and
  Cosmic Microwave Background}},
\newblock \bibinfo{journal}{PoS} \bibinfo{volume}{EDSU2018}
  (\bibinfo{year}{2018}) \bibinfo{pages}{037}.
  \DOIprefix\doi{10.22323/1.335.0037}.
  \href{http://arxiv.org/abs/1901.05289}{{\tt arXiv:1901.05289}}.
\bibitem[{Box(1976)}]{Box1976}
\bibinfo{author}{G.~E.~P. Box},
\newblock \bibinfo{title}{Science and statistics},
\newblock \bibinfo{journal}{Journal of the American Statistical Association}
  \bibinfo{volume}{71} (\bibinfo{year}{1976}) \bibinfo{pages}{791--799}.
  \DOIprefix\doi{10.1080/01621459.1976.10480949}.
\bibitem[{Wands et~al.(2012)Wands, De-Santiago, and Wang}]{Wands:2012vg}
\bibinfo{author}{D.~Wands}, \bibinfo{author}{J.~De-Santiago},
  \bibinfo{author}{Y.~Wang},
\newblock \bibinfo{title}{{Inhomogeneous vacuum energy}},
\newblock \bibinfo{journal}{Classical and Quantum Gravity} \bibinfo{volume}{29}
  (\bibinfo{year}{2012}) \bibinfo{pages}{145017}.
  \DOIprefix\doi{10.1088/0264-9381/29/14/145017}.
  \href{http://arxiv.org/abs/1203.6776}{{\tt arXiv:1203.6776}}.
\bibitem[{Salvatelli et~al.(2014)Salvatelli, Said, Bruni, Melchiorri, and
  Wands}]{Salvatelli2014}
\bibinfo{author}{V.~Salvatelli}, \bibinfo{author}{N.~Said},
  \bibinfo{author}{M.~Bruni}, \bibinfo{author}{A.~Melchiorri},
  \bibinfo{author}{D.~Wands},
\newblock \bibinfo{title}{{Indications of a Late-Time Interaction in the Dark
  Sector}},
\newblock \bibinfo{journal}{Physical Review Letters} \bibinfo{volume}{113}
  (\bibinfo{year}{2014}) \bibinfo{pages}{181301}.
  \DOIprefix\doi{10.1103/PhysRevLett.113.181301}.
\bibitem[{Sol\`{a} et~al.(2017)Sol\`{a}, G\'{o}mez-Valent, and
  de~Cruz~P\'{e}rez}]{Sola:2017lxc}
\bibinfo{author}{J.~Sol\`{a}}, \bibinfo{author}{A.~G\'{o}mez-Valent},
  \bibinfo{author}{J.~de~Cruz~P\'{e}rez},
\newblock \bibinfo{title}{{Vacuum dynamics in the Universe versus a rigid
  $\Lambda=$const}},
\newblock \bibinfo{journal}{International Journal of Modern Physics}
  \bibinfo{volume}{A32} (\bibinfo{year}{2017}) \bibinfo{pages}{1730014}.
  \DOIprefix\doi{10.1142/S0217751X17300149}.
  \href{http://arxiv.org/abs/1709.07451}{{\tt arXiv:1709.07451}}.
\bibitem[{Kumar and Nunes(2017)}]{Kumar:2017dnp}
\bibinfo{author}{S.~Kumar}, \bibinfo{author}{R.~C. Nunes},
\newblock \bibinfo{title}{{Echo of interactions in the dark sector}},
\newblock \bibinfo{journal}{Phys. Rev.} \bibinfo{volume}{D96}
  (\bibinfo{year}{2017}) \bibinfo{pages}{103511}.
  \DOIprefix\doi{10.1103/PhysRevD.96.103511}.
  \href{http://arxiv.org/abs/1702.02143}{{\tt arXiv:1702.02143}}.
\bibitem[{{Martinelli} et~al.(2019){Martinelli}, {Hogg}, {Peirone}, {Bruni},
  and {Wands}}]{Martinelli:2019dau}
\bibinfo{author}{M.~{Martinelli}}, \bibinfo{author}{N.~B. {Hogg}},
  \bibinfo{author}{S.~{Peirone}}, \bibinfo{author}{M.~{Bruni}},
  \bibinfo{author}{D.~{Wands}},
\newblock \bibinfo{title}{{Constraints on the interacting vacuum-geodesic CDM
  scenario}},
\newblock \bibinfo{journal}{Monthly Notices of the Royal Astronomical Society}
  \bibinfo{volume}{488} (\bibinfo{year}{2019}) \bibinfo{pages}{3423--3438}.
  \DOIprefix\doi{10.1093/mnras/stz1915}.
  \href{http://arxiv.org/abs/1902.10694}{{\tt arXiv:1902.10694}}.
\bibitem[{{Pan} et~al.(2019){Pan}, {Yang}, {Di Valentino}, {Saridakis}, and
  {Chakraborty}}]{Pan2019}
\bibinfo{author}{S.~{Pan}}, \bibinfo{author}{W.~{Yang}},
  \bibinfo{author}{E.~{Di Valentino}}, \bibinfo{author}{E.~N. {Saridakis}},
  \bibinfo{author}{S.~{Chakraborty}},
\newblock \bibinfo{title}{{Interacting scenarios with dynamical dark energy:
  observational constraints and alleviation of the $H_0$ tension}},
\newblock \bibinfo{journal}{ArXiv e-prints}  (\bibinfo{year}{2019}).
  \href{http://arxiv.org/abs/1907.07540}{{\tt arXiv:1907.07540}}.
\bibitem[{Pan et~al.(2020)Pan, Sharov, and Yang}]{Pan:2020zza}
\bibinfo{author}{S.~Pan}, \bibinfo{author}{G.~S. Sharov},
  \bibinfo{author}{W.~Yang},
\newblock \bibinfo{title}{{Field theoretic interpretations of interacting dark
  energy scenarios and recent observations}}  (\bibinfo{year}{2020}).
  \href{http://arxiv.org/abs/2001.03120}{{\tt arXiv:2001.03120}}.
\bibitem[{{Copeland} et~al.(2006){Copeland}, {Sami}, and
  {Tsujikawa}}]{Copeland2006}
\bibinfo{author}{E.~J. {Copeland}}, \bibinfo{author}{M.~{Sami}},
  \bibinfo{author}{S.~{Tsujikawa}},
\newblock \bibinfo{title}{{Dynamics of Dark Energy}},
\newblock \bibinfo{journal}{International Journal of Modern Physics D}
  \bibinfo{volume}{15} (\bibinfo{year}{2006}) \bibinfo{pages}{1753--1935}.
  \DOIprefix\doi{10.1142/S021827180600942X}.
  \href{http://arxiv.org/abs/hep-th/0603057}{{\tt arXiv:hep-th/0603057}}.
\bibitem[{{Clifton} et~al.(2012){Clifton}, {Ferreira}, {Padilla}, and
  {Skordis}}]{Clifton2012}
\bibinfo{author}{T.~{Clifton}}, \bibinfo{author}{P.~G. {Ferreira}},
  \bibinfo{author}{A.~{Padilla}}, \bibinfo{author}{C.~{Skordis}},
\newblock \bibinfo{title}{{Modified gravity and cosmology}},
\newblock \bibinfo{journal}{Physics Reports} \bibinfo{volume}{513}
  (\bibinfo{year}{2012}) \bibinfo{pages}{1--189}.
  \DOIprefix\doi{10.1016/j.physrep.2012.01.001}.
  \href{http://arxiv.org/abs/1106.2476}{{\tt arXiv:1106.2476}}.
\bibitem[{{Joyce} et~al.(2016){Joyce}, {Lombriser}, and {Schmidt}}]{Joyce2016}
\bibinfo{author}{A.~{Joyce}}, \bibinfo{author}{L.~{Lombriser}},
  \bibinfo{author}{F.~{Schmidt}},
\newblock \bibinfo{title}{{Dark Energy Versus Modified Gravity}},
\newblock \bibinfo{journal}{Annual Review of Nuclear and Particle Science}
  \bibinfo{volume}{66} (\bibinfo{year}{2016}) \bibinfo{pages}{95--122}.
  \DOIprefix\doi{10.1146/annurev-nucl-102115-044553}.
  \href{http://arxiv.org/abs/1601.06133}{{\tt arXiv:1601.06133}}.
\bibitem[{{Ezquiaga} and {Zumalac{\'a}rregui}(2017)}]{Ezquiaga2017}
\bibinfo{author}{J.~M. {Ezquiaga}}, \bibinfo{author}{M.~{Zumalac{\'a}rregui}},
\newblock \bibinfo{title}{{Dark Energy After GW170817: Dead Ends and the Road
  Ahead}},
\newblock \bibinfo{journal}{Physical Review Letters} \bibinfo{volume}{119}
  (\bibinfo{year}{2017}) \bibinfo{pages}{251304}.
  \DOIprefix\doi{10.1103/PhysRevLett.119.251304}.
  \href{http://arxiv.org/abs/1710.05901}{{\tt arXiv:1710.05901}}.
\bibitem[{Frusciante and Perenon(2019)}]{Frusciante:2019xia}
\bibinfo{author}{N.~Frusciante}, \bibinfo{author}{L.~Perenon},
\newblock \bibinfo{title}{{Effective Field Theory of Dark Energy: a Review}}
  (\bibinfo{year}{2019}). \href{http://arxiv.org/abs/1907.03150}{{\tt
  arXiv:1907.03150}}.
\bibitem[{Lewis et~al.(2000)Lewis, Challinor, and Lasenby}]{Lewis:1999bs}
\bibinfo{author}{A.~Lewis}, \bibinfo{author}{A.~Challinor},
  \bibinfo{author}{A.~Lasenby},
\newblock \bibinfo{title}{{Efficient computation of CMB anisotropies in closed
  FRW models}},
\newblock \bibinfo{journal}{The Astrophysical Journal} \bibinfo{volume}{538}
  (\bibinfo{year}{2000}) \bibinfo{pages}{473--476}.
  \DOIprefix\doi{10.1086/309179}.
  \href{http://arxiv.org/abs/astro-ph/9911177}{{\tt arXiv:astro-ph/9911177}}.
\bibitem[{Howlett et~al.(2012)Howlett, Lewis, Hall, and
  Challinor}]{Howlett:2012mh}
\bibinfo{author}{C.~Howlett}, \bibinfo{author}{A.~Lewis},
  \bibinfo{author}{A.~Hall}, \bibinfo{author}{A.~Challinor},
\newblock \bibinfo{title}{{CMB power spectrum parameter degeneracies in the era
  of precision cosmology}},
\newblock \bibinfo{journal}{Journal of Cosmology and Astroparticle Physics}
  \bibinfo{volume}{1204} (\bibinfo{year}{2012}) \bibinfo{pages}{027}.
  \DOIprefix\doi{10.1088/1475-7516/2012/04/027}.
  \href{http://arxiv.org/abs/1201.3654}{{\tt arXiv:1201.3654}}.
\bibitem[{Lewis and Bridle(2002)}]{Lewis:2002ah}
\bibinfo{author}{A.~Lewis}, \bibinfo{author}{S.~Bridle},
\newblock \bibinfo{title}{{Cosmological parameters from CMB and other data: A
  Monte Carlo approach}},
\newblock \bibinfo{journal}{Physical Review D} \bibinfo{volume}{66}
  (\bibinfo{year}{2002}) \bibinfo{pages}{103511}.
  \DOIprefix\doi{10.1103/PhysRevD.66.103511}.
  \href{http://arxiv.org/abs/astro-ph/0205436}{{\tt arXiv:astro-ph/0205436}}.
\bibitem[{Lewis(2013)}]{Lewis:2013hha}
\bibinfo{author}{A.~Lewis},
\newblock \bibinfo{title}{{Efficient sampling of fast and slow cosmological
  parameters}},
\newblock \bibinfo{journal}{Physical Review D} \bibinfo{volume}{87}
  (\bibinfo{year}{2013}) \bibinfo{pages}{103529}.
  \DOIprefix\doi{10.1103/PhysRevD.87.103529}.
  \href{http://arxiv.org/abs/1304.4473}{{\tt arXiv:1304.4473}}.
\bibitem[{Crittenden et~al.(2009)Crittenden, Pogosian, and
  Zhao}]{Crittenden:2005wj}
\bibinfo{author}{R.~G. Crittenden}, \bibinfo{author}{L.~Pogosian},
  \bibinfo{author}{G.-B. Zhao},
\newblock \bibinfo{title}{{Investigating dark energy experiments with principal
  components}},
\newblock \bibinfo{journal}{Journal of Cosmology and Astroparticle Physics}
  \bibinfo{volume}{0912} (\bibinfo{year}{2009}) \bibinfo{pages}{025}.
  \DOIprefix\doi{10.1088/1475-7516/2009/12/025}.
  \href{http://arxiv.org/abs/astro-ph/0510293}{{\tt arXiv:astro-ph/0510293}}.
\bibitem[{{Crittenden} et~al.(2012){Crittenden}, {Zhao}, {Pogosian},
  {Samushia}, and {Zhang}}]{Crittenden2012}
\bibinfo{author}{R.~G. {Crittenden}}, \bibinfo{author}{G.-B. {Zhao}},
  \bibinfo{author}{L.~{Pogosian}}, \bibinfo{author}{L.~{Samushia}},
  \bibinfo{author}{X.~{Zhang}},
\newblock \bibinfo{title}{{Fables of reconstruction: controlling bias in the
  dark energy equation of state}},
\newblock \bibinfo{journal}{Journal of Cosmology and Astroparticle Physics}
  \bibinfo{volume}{2} (\bibinfo{year}{2012}) \bibinfo{pages}{048}.
  \DOIprefix\doi{10.1088/1475-7516/2012/02/048}.
  \href{http://arxiv.org/abs/1112.1693}{{\tt arXiv:1112.1693}}.
\bibitem[{Wang et~al.(2018)Wang, Pogosian, Zhao, and Zucca}]{Wang:2018fng}
\bibinfo{author}{Y.~Wang}, \bibinfo{author}{L.~Pogosian},
  \bibinfo{author}{G.-B. Zhao}, \bibinfo{author}{A.~Zucca},
\newblock \bibinfo{title}{{Evolution of dark energy reconstructed from the
  latest observations}},
\newblock \bibinfo{journal}{Astrophys. J.} \bibinfo{volume}{869}
  (\bibinfo{year}{2018}) \bibinfo{pages}{L8}.
  \DOIprefix\doi{10.3847/2041-8213/aaf238}.
  \href{http://arxiv.org/abs/1807.03772}{{\tt arXiv:1807.03772}}.
\bibitem[{Gerardi et~al.(2019)Gerardi, Martinelli, and
  Silvestri}]{Gerardi:2019obr}
\bibinfo{author}{F.~Gerardi}, \bibinfo{author}{M.~Martinelli},
  \bibinfo{author}{A.~Silvestri},
\newblock \bibinfo{title}{{Reconstruction of the Dark Energy equation of state
  from latest data: the impact of theoretical priors}},
\newblock \bibinfo{journal}{JCAP} \bibinfo{volume}{1907} (\bibinfo{year}{2019})
  \bibinfo{pages}{042}. \DOIprefix\doi{10.1088/1475-7516/2019/07/042}.
  \href{http://arxiv.org/abs/1902.09423}{{\tt arXiv:1902.09423}}.
\bibitem[{Dam et~al.(2019)Dam, Bolejko, and Lewis}]{Dam:2019prv}
\bibinfo{author}{L.~Dam}, \bibinfo{author}{K.~Bolejko}, \bibinfo{author}{G.~F.
  Lewis},
\newblock \bibinfo{title}{{Probing the independence within the dark sector in
  the fluid approximation}},
\newblock \bibinfo{journal}{JCAP} \bibinfo{volume}{1912} (\bibinfo{year}{2019})
  \bibinfo{pages}{030}. \DOIprefix\doi{10.1088/1475-7516/2019/12/030}.
  \href{http://arxiv.org/abs/1908.01953}{{\tt arXiv:1908.01953}}.
\bibitem[{Wang et~al.(2015)Wang, Zhao, Wands, Pogosian, and
  Crittenden}]{Wang:2015wga}
\bibinfo{author}{Y.~Wang}, \bibinfo{author}{G.-B. Zhao},
  \bibinfo{author}{D.~Wands}, \bibinfo{author}{L.~Pogosian},
  \bibinfo{author}{R.~G. Crittenden},
\newblock \bibinfo{title}{{Reconstruction of the dark matter--vacuum energy
  interaction}},
\newblock \bibinfo{journal}{Physical Review D} \bibinfo{volume}{92}
  (\bibinfo{year}{2015}) \bibinfo{pages}{103005}.
  \DOIprefix\doi{10.1103/PhysRevD.92.103005}.
  \href{http://arxiv.org/abs/1505.01373}{{\tt arXiv:1505.01373}}.
\bibitem[{Wang et~al.(2014)Wang, Wands, Zhao, and Xu}]{Wang:2014xca}
\bibinfo{author}{Y.~Wang}, \bibinfo{author}{D.~Wands}, \bibinfo{author}{G.-B.
  Zhao}, \bibinfo{author}{L.~Xu},
\newblock \bibinfo{title}{{Post-$Planck$ constraints on interacting vacuum
  energy}},
\newblock \bibinfo{journal}{Phys. Rev.} \bibinfo{volume}{D90}
  (\bibinfo{year}{2014}) \bibinfo{pages}{023502}.
  \DOIprefix\doi{10.1103/PhysRevD.90.023502}.
  \href{http://arxiv.org/abs/1404.5706}{{\tt arXiv:1404.5706}}.
\bibitem[{{Beutler} et~al.(2011){Beutler}, {Blake}, {Colless}, {Jones},
  {Staveley-Smith}, {Campbell}, {Parker}, {Saunders}, and
  {Watson}}]{Beutler2011}
\bibinfo{author}{F.~{Beutler}}, \bibinfo{author}{C.~{Blake}},
  \bibinfo{author}{M.~{Colless}}, \bibinfo{author}{D.~H. {Jones}},
  \bibinfo{author}{L.~{Staveley-Smith}}, \bibinfo{author}{L.~{Campbell}},
  \bibinfo{author}{Q.~{Parker}}, \bibinfo{author}{W.~{Saunders}},
  \bibinfo{author}{F.~{Watson}},
\newblock \bibinfo{title}{{The 6dF Galaxy Survey: baryon acoustic oscillations
  and the local Hubble constant}},
\newblock \bibinfo{journal}{Monthly Notices of the Royal Astronomical Society}
  \bibinfo{volume}{416} (\bibinfo{year}{2011}) \bibinfo{pages}{3017--3032}.
  \DOIprefix\doi{10.1111/j.1365-2966.2011.19250.x}.
  \href{http://arxiv.org/abs/1106.3366}{{\tt arXiv:1106.3366}}.
\bibitem[{Alam et~al.(2017)}]{Alam:2016hwk}
\bibinfo{author}{S.~Alam}, et~al. (\bibinfo{collaboration}{BOSS}),
\newblock \bibinfo{title}{{The clustering of galaxies in the completed SDSS-III
  Baryon Oscillation Spectroscopic Survey: cosmological analysis of the DR12
  galaxy sample}},
\newblock \bibinfo{journal}{Monthly Notices of the Royal Astronomical Society}
  \bibinfo{volume}{470} (\bibinfo{year}{2017}) \bibinfo{pages}{2617--2652}.
  \DOIprefix\doi{10.1093/mnras/stx721}.
  \href{http://arxiv.org/abs/1607.03155}{{\tt arXiv:1607.03155}}.
\bibitem[{Scolnic et~al.(2018)}]{Scolnic:2017caz}
\bibinfo{author}{D.~M. Scolnic}, et~al.,
\newblock \bibinfo{title}{{The Complete Light-curve Sample of Spectroscopically
  Confirmed SNe Ia from Pan-STARRS1 and Cosmological Constraints from the
  Combined Pantheon Sample}},
\newblock \bibinfo{journal}{Astrophys. J.} \bibinfo{volume}{859}
  (\bibinfo{year}{2018}) \bibinfo{pages}{101}.
  \DOIprefix\doi{10.3847/1538-4357/aab9bb}.
  \href{http://arxiv.org/abs/1710.00845}{{\tt arXiv:1710.00845}}.
\bibitem[{{Di Valentino} et~al.(2019){Di Valentino}, {Melchiorri}, {Mena}, and
  {Vagnozzi}}]{diValentino2019}
\bibinfo{author}{E.~{Di Valentino}}, \bibinfo{author}{A.~{Melchiorri}},
  \bibinfo{author}{O.~{Mena}}, \bibinfo{author}{S.~{Vagnozzi}},
\newblock \bibinfo{title}{{Interacting dark energy after the latest Planck,
  DES, and $H_0$ measurements: an excellent solution to the $H_0$ and cosmic
  shear tensions}},
\newblock \bibinfo{journal}{arXiv e-prints}  (\bibinfo{year}{2019}).
  \href{http://arxiv.org/abs/1908.04281}{{\tt arXiv:1908.04281}}.
\bibitem[{Poulin et~al.(2018)Poulin, Boddy, Bird, and
  Kamionkowski}]{Poulin:2018zxs}
\bibinfo{author}{V.~Poulin}, \bibinfo{author}{K.~K. Boddy},
  \bibinfo{author}{S.~Bird}, \bibinfo{author}{M.~Kamionkowski},
\newblock \bibinfo{title}{{Implications of an extended dark energy cosmology
  with massive neutrinos for cosmological tensions}},
\newblock \bibinfo{journal}{Phys. Rev.} \bibinfo{volume}{D97}
  (\bibinfo{year}{2018}) \bibinfo{pages}{123504}.
  \DOIprefix\doi{10.1103/PhysRevD.97.123504}.
  \href{http://arxiv.org/abs/1803.02474}{{\tt arXiv:1803.02474}}.
\bibitem[{Martinelli and Tutusaus(2019)}]{Martinelli:2019krf}
\bibinfo{author}{M.~Martinelli}, \bibinfo{author}{I.~Tutusaus},
\newblock \bibinfo{title}{{CMB tensions with low-redshift $H_0$ and $S_8$
  measurements: impact of a redshift-dependent type-Ia supernovae intrinsic
  luminosity}},
\newblock \bibinfo{journal}{Symmetry} \bibinfo{volume}{11}
  (\bibinfo{year}{2019}) \bibinfo{pages}{986}.
  \DOIprefix\doi{10.3390/sym11080986}.
  \href{http://arxiv.org/abs/1906.09189}{{\tt arXiv:1906.09189}}.
\bibitem[{Borges and Wands(2017)}]{Borges:2017jvi}
\bibinfo{author}{H.~A. Borges}, \bibinfo{author}{D.~Wands},
\newblock \bibinfo{title}{{Growth of structure in interacting vacuum
  cosmologies}}  (\bibinfo{year}{2017}).
  \href{http://arxiv.org/abs/1709.08933}{{\tt arXiv:1709.08933}}.
\bibitem[{Aylor et~al.(2019)Aylor, Joy, Knox, Millea, Raghunathan, and
  Wu}]{Aylor:2018drw}
\bibinfo{author}{K.~Aylor}, \bibinfo{author}{M.~Joy},
  \bibinfo{author}{L.~Knox}, \bibinfo{author}{M.~Millea},
  \bibinfo{author}{S.~Raghunathan}, \bibinfo{author}{W.~L.~K. Wu},
\newblock \bibinfo{title}{{Sounds Discordant: Classical Distance Ladder \&
  $\Lambda$CDM -based Determinations of the Cosmological Sound Horizon}},
\newblock \bibinfo{journal}{Astrophys. J.} \bibinfo{volume}{874}
  (\bibinfo{year}{2019}) \bibinfo{pages}{4}.
  \DOIprefix\doi{10.3847/1538-4357/ab0898}.
  \href{http://arxiv.org/abs/1811.00537}{{\tt arXiv:1811.00537}}.
\bibitem[{Arendse et~al.(2019)}]{Arendse:2019hev}
\bibinfo{author}{N.~Arendse}, et~al.,
\newblock \bibinfo{title}{{Cosmic dissonance: new physics or systematics behind
  a short sound horizon?}}  (\bibinfo{year}{2019}).
  \href{http://arxiv.org/abs/1909.07986}{{\tt arXiv:1909.07986}}.
\bibitem[{Knox and Millea(2019)}]{Knox:2019rjx}
\bibinfo{author}{L.~Knox}, \bibinfo{author}{M.~Millea},
\newblock \bibinfo{title}{{The Hubble Hunter's Guide}}  (\bibinfo{year}{2019}).
  \href{http://arxiv.org/abs/1908.03663}{{\tt arXiv:1908.03663}}.
\bibitem[{Abbott et~al.(2018)}]{Abbott:2017wau}
\bibinfo{author}{T.~Abbott}, et~al. (\bibinfo{collaboration}{DES}),
\newblock \bibinfo{title}{{Dark Energy Survey year 1 results: Cosmological
  constraints from galaxy clustering and weak lensing}},
\newblock \bibinfo{journal}{Phys. Rev. D} \bibinfo{volume}{98}
  (\bibinfo{year}{2018}) \bibinfo{pages}{043526}.
  \DOIprefix\doi{10.1103/PhysRevD.98.043526}.
  \href{http://arxiv.org/abs/1708.01530}{{\tt arXiv:1708.01530}}.
\bibitem[{Rasmussen and Williams(2006)}]{Rasmussen}
\bibinfo{author}{C.~E. Rasmussen}, \bibinfo{author}{C.~K.~I. Williams},
  \bibinfo{title}{Gaussian Processes for Machine Learning},
  \bibinfo{publisher}{MIT Press}, \bibinfo{year}{2006}.
\bibitem[{Ambikasaran et~al.(2015)Ambikasaran, {Foreman-Mackey}, {Greengard},
  {Hogg}, and {O'Neil}}]{Ambikasaran2015}
\bibinfo{author}{S.~Ambikasaran}, \bibinfo{author}{D.~{Foreman-Mackey}},
  \bibinfo{author}{L.~{Greengard}}, \bibinfo{author}{D.~W. {Hogg}},
  \bibinfo{author}{M.~{O'Neil}},
\newblock \bibinfo{title}{{Fast Direct Methods for Gaussian Processes}},
\newblock \bibinfo{journal}{IEEE Transactions on Pattern Analysis and Machine
  Intelligence} \bibinfo{volume}{38} (\bibinfo{year}{2015}).
  \DOIprefix\doi{10.1109/TPAMI.2015.2448083}.
  \href{http://arxiv.org/abs/1403.6015}{{\tt arXiv:1403.6015}}.
\bibitem[{Croft et~al.(2018)Croft, Romeo, and Metcalf}]{Croft:2017tur}
\bibinfo{author}{R.~A.~C. Croft}, \bibinfo{author}{A.~Romeo},
  \bibinfo{author}{R.~B. Metcalf},
\newblock \bibinfo{title}{{Weak lensing of the Lyman $\boldsymbol {\alpha }$
  forest}},
\newblock \bibinfo{journal}{Monthly Notices of the Royal Astronomical Society}
  \bibinfo{volume}{477} (\bibinfo{year}{2018}) \bibinfo{pages}{1814--1821}.
  \DOIprefix\doi{10.1093/mnras/sty650}.
  \href{http://arxiv.org/abs/1706.07870}{{\tt arXiv:1706.07870}}.
\bibitem[{{Braun} et~al.(2015){Braun}, {Bourke}, {Green}, {Keane}, and
  {Wagg}}]{Braun2015}
\bibinfo{author}{R.~{Braun}}, \bibinfo{author}{T.~{Bourke}},
  \bibinfo{author}{J.~A. {Green}}, \bibinfo{author}{E.~{Keane}},
  \bibinfo{author}{J.~{Wagg}},
\newblock \bibinfo{title}{{Advancing Astrophysics with the Square Kilometre
  Array}},
\newblock \bibinfo{journal}{Proceedings of Science (AASKA14)}
  (\bibinfo{year}{2015}) \bibinfo{pages}{174}.
\bibitem[{Bacon et~al.(2018)}]{SKA2018}
\bibinfo{author}{D.~J. Bacon}, et~al.,
\newblock \bibinfo{title}{{Cosmology with Phase 1 of the Square Kilometre
  Array; Red Book 2018: Technical specifications and performance forecasts}},
\newblock \bibinfo{journal}{arXiv e-prints}  (\bibinfo{year}{2018})
  \bibinfo{pages}{arXiv:1811.02743}.
  \href{http://arxiv.org/abs/1811.02743}{{\tt arXiv:1811.02743}}.
\bibitem[{Huterer and Starkman(2003)}]{Huterer:2002hy}
\bibinfo{author}{D.~Huterer}, \bibinfo{author}{G.~Starkman},
\newblock \bibinfo{title}{{Parameterization of dark-energy properties: A
  Principal-component approach}},
\newblock \bibinfo{journal}{Phys. Rev. Lett.} \bibinfo{volume}{90}
  (\bibinfo{year}{2003}) \bibinfo{pages}{031301}.
  \DOIprefix\doi{10.1103/PhysRevLett.90.031301}.
  \href{http://arxiv.org/abs/astro-ph/0207517}{{\tt arXiv:astro-ph/0207517}}.
\bibitem[{Lewis(2019)}]{Lewis:2019xzd}
\bibinfo{author}{A.~Lewis},
\newblock \bibinfo{title}{{GetDist: a Python package for analysing Monte Carlo
  samples}}  (\bibinfo{year}{2019}). \URLprefix
  \url{https://getdist.readthedocs.io}.
  \href{http://arxiv.org/abs/1910.13970}{{\tt arXiv:1910.13970}}.
\bibitem[{Bayes(1763)}]{Bayes1763}
\bibinfo{author}{T.~Bayes},
\newblock \bibinfo{title}{{LII. An essay towards solving a problem in the
  doctrine of chances. By the late Rev. Mr. Bayes, F. R. S. communicated by Mr.
  Price, in a letter to John Canton, A. M. F. R. S}},
\newblock \bibinfo{journal}{Philosophical Transactions of the Royal Society of
  London} \bibinfo{volume}{53} (\bibinfo{year}{1763})
  \bibinfo{pages}{370--418}. \DOIprefix\doi{10.1098/rstl.1763.0053}.
\bibitem[{{Linares Cede{\~n}o} et~al.(2019){Linares Cede{\~n}o}, {Montiel},
  {Hidalgo}, and {Germ{\'a}n}}]{Linares2019}
\bibinfo{author}{F.~X. {Linares Cede{\~n}o}}, \bibinfo{author}{A.~{Montiel}},
  \bibinfo{author}{J.~C. {Hidalgo}}, \bibinfo{author}{G.~{Germ{\'a}n}},
\newblock \bibinfo{title}{{Bayesian evidence for $\alpha$-attractor dark energy
  models}},
\newblock \bibinfo{journal}{JCAP} \bibinfo{volume}{1908} (\bibinfo{year}{2019})
  \bibinfo{pages}{002}. \DOIprefix\doi{10.1088/1475-7516/2019/08/002}.
  \href{http://arxiv.org/abs/1905.00834}{{\tt arXiv:1905.00834}}.
\bibitem[{{Heavens} et~al.(2017){Heavens}, {Fantaye}, {Mootoovaloo}, {Eggers},
  {Hosenie}, {Kroon}, and {Sellentin}}]{Heavens:2017afc}
\bibinfo{author}{A.~{Heavens}}, \bibinfo{author}{Y.~{Fantaye}},
  \bibinfo{author}{A.~{Mootoovaloo}}, \bibinfo{author}{H.~{Eggers}},
  \bibinfo{author}{Z.~{Hosenie}}, \bibinfo{author}{S.~{Kroon}},
  \bibinfo{author}{E.~{Sellentin}},
\newblock \bibinfo{title}{{Marginal Likelihoods from Monte Carlo Markov
  Chains}},
\newblock \bibinfo{journal}{ArXiv e-prints}  (\bibinfo{year}{2017}).
  \href{http://arxiv.org/abs/1704.03472}{{\tt arXiv:1704.03472}}.
\bibitem[{Hobson et~al.(2010)Hobson, Jaffe, Liddle, Mukherjee, and
  Parkinson}]{Hobson}
\bibinfo{author}{M.~P. Hobson}, \bibinfo{author}{A.~H. Jaffe},
  \bibinfo{author}{A.~R. Liddle}, \bibinfo{author}{P.~Mukherjee},
  \bibinfo{author}{D.~Parkinson}, \bibinfo{title}{Bayesian Methods in
  Cosmology}, \bibinfo{publisher}{Cambridge University Press},
  \bibinfo{year}{2010}.
\bibitem[{Jeffreys(1961)}]{Jeffreys}
\bibinfo{author}{H.~Jeffreys}, \bibinfo{title}{Theory of Probability},
  \bibinfo{publisher}{Clarendon Press, Oxford}, \bibinfo{year}{1961}.
\bibitem[{{Efstathiou}(2008)}]{Efstathiou2008}
\bibinfo{author}{G.~{Efstathiou}},
\newblock \bibinfo{title}{{Limitations of Bayesian Evidence applied to
  cosmology}},
\newblock \bibinfo{journal}{Monthly Notices of the Royal Astronomical Society}
  \bibinfo{volume}{388} (\bibinfo{year}{2008}) \bibinfo{pages}{1314--1320}.
  \DOIprefix\doi{10.1111/j.1365-2966.2008.13498.x}.
  \href{http://arxiv.org/abs/0802.3185}{{\tt arXiv:0802.3185}}.
\bibitem[{Zhao et~al.(2017)}]{Zhao:2017cud}
\bibinfo{author}{G.-B. Zhao}, et~al.,
\newblock \bibinfo{title}{{Dynamical dark energy in light of the latest
  observations}},
\newblock \bibinfo{journal}{Nature Astronomy} \bibinfo{volume}{1}
  (\bibinfo{year}{2017}) \bibinfo{pages}{627--632}.
  \DOIprefix\doi{10.1038/s41550-017-0216-z}.
  \href{http://arxiv.org/abs/1701.08165}{{\tt arXiv:1701.08165}}.
\bibitem[{Yang et~al.(2019{\natexlab{a}})Yang, Pan, Di~Valentino, Saridakis,
  and Chakraborty}]{Yang:2018qmz}
\bibinfo{author}{W.~Yang}, \bibinfo{author}{S.~Pan},
  \bibinfo{author}{E.~Di~Valentino}, \bibinfo{author}{E.~N. Saridakis},
  \bibinfo{author}{S.~Chakraborty},
\newblock \bibinfo{title}{{Observational constraints on one-parameter dynamical
  dark-energy parametrizations and the $H_0$ tension}},
\newblock \bibinfo{journal}{Phys. Rev.} \bibinfo{volume}{D99}
  (\bibinfo{year}{2019}{\natexlab{a}}) \bibinfo{pages}{043543}.
  \DOIprefix\doi{10.1103/PhysRevD.99.043543}.
  \href{http://arxiv.org/abs/1810.05141}{{\tt arXiv:1810.05141}}.
\bibitem[{Yang et~al.(2019{\natexlab{b}})Yang, Banerjee, Paliathanasis, and
  Pan}]{Yang:2018qec}
\bibinfo{author}{W.~Yang}, \bibinfo{author}{N.~Banerjee},
  \bibinfo{author}{A.~Paliathanasis}, \bibinfo{author}{S.~Pan},
\newblock \bibinfo{title}{{Reconstructing the dark matter and dark energy
  interaction scenarios from observations}},
\newblock \bibinfo{journal}{Phys. Dark Univ.} \bibinfo{volume}{26}
  (\bibinfo{year}{2019}{\natexlab{b}}) \bibinfo{pages}{100383}.
  \DOIprefix\doi{10.1016/j.dark.2019.100383}.
  \href{http://arxiv.org/abs/1812.06854}{{\tt arXiv:1812.06854}}.
\bibitem[{Lucca and Hooper(2020)}]{Lucca:2020zjb}
\bibinfo{author}{M.~Lucca}, \bibinfo{author}{D.~C. Hooper},
\newblock \bibinfo{title}{{Tensions in the dark: shedding light on Dark
  Matter-Dark Energy Interactions}}  (\bibinfo{year}{2020}).
  \href{http://arxiv.org/abs/2002.06127}{{\tt arXiv:2002.06127}}.

\end{thebibliography}

\end{document}